\documentclass[a4paper,12pt]{article}

\usepackage{mathrsfs,graphicx,rotating,amsmath,amsfonts,mathtools,booktabs,amssymb}
\usepackage{hyperref}\usepackage{slashed}
\usepackage[nosort]{cite}
\usepackage[table,xcdraw]{xcolor}
\usepackage{youngtab}
\usepackage{marginnote}
\usepackage{soul}
\usepackage{latexsym}

\usepackage{graphicx}
\usepackage{multirow,multicol}
\hypersetup{colorlinks,bookmarksopen,bookmarksnumbered,
linkcolor=blus,pdfstartview=FitH,urlcolor=rossos,citecolor=verde}
\allowdisplaybreaks

\newcommand{\GF}{G_{\rm F}}
\newcommand{\GN}{G_{\rm N}}
\newcommand{\mio}[1]{}

 \newcommand{\fig}[1]{~\ref{fig:#1}}
\newcommand{\sfrac}[2]{#1/#2}

\allowdisplaybreaks
\usepackage{multicol}
\usepackage{color}
\definecolor{darkgreen}{rgb}{0,0.5,0}
\definecolor{rosso}{cmyk}{0,1,1,0.4}
\definecolor{rossos}{cmyk}{0,1,1,0.55}
\definecolor{rossoc}{cmyk}{0,1,1,0.2}
\definecolor{blu}{cmyk}{1,1,0,0.3}
\definecolor{blus}{cmyk}{1,1,0,0.6}
\definecolor{bluc}{cmyk}{1,1,0,0.1}
\definecolor{verde}{cmyk}{0.92,0,0.59,0.25}
\definecolor{verdec}{cmyk}{0.92,0,0.59,0.15}
\definecolor{verdes}{cmyk}{0.92,0,0.59,0.4}
\definecolor{beaublue}{rgb}{0.74, 0.83, 0.9}
\definecolor{columbiablue}{rgb}{0.61, 0.87, 1.0}
\definecolor{navyblue}{rgb}{0.0, 0.0, 0.5}

\newcommand{\eq}[1]{~{\rm (\ref{eq:#1})}}

\newcommand{\keV}{\,{\rm keV}}
\newcommand{\MeV}{\,{\rm MeV}}
\newcommand{\GeV}{\,{\rm GeV}}

\def\circa#1{\,\raise.3ex\hbox{$#1$\kern-.75em\lower1ex\hbox{$\sim$}}\,}

\newcommand{\beq}{\begin{equation}}
\newcommand{\eeq}{\end{equation}}

\newcommand{\bea}{\begin{eqnarray}}
\newcommand{\eea}{\end{eqnarray}}
\newcommand{\be}{\begin{equation}}
\newcommand{\ee}{\end{equation}}
\font\tenrsfs=rsfs10 at 12pt
\font\sevenrsfs=rsfs7
\font\fiversfs=rsfs5
\newfam\rsfsfam
\textfont\rsfsfam=\tenrsfs
\scriptfont\rsfsfam=\sevenrsfs
\scriptscriptfont\rsfsfam=\fiversfs

\newsavebox\MBox

\def\circa#1{\,\raise.3ex\hbox{$#1$\kern-.75em\lower1ex\hbox{$\sim$}}\,}
\makeatletter

\font\ital=cmu10 

\def\hhref#1{\href{http://arxiv.org/abs/#1}{arXiv:#1}}
\usepackage{xstring} 
\newcommand{\hhrefq}[1]{\IfSubStr{#1}{:}{\href{http://inspirehep.net/search?ln=en&ln=en&p=#1&of=hb&action_search=Search&sf=&so=d&rm=&rg=25&sc=0}{InSpire:#1}}{\hhref{#1}}}

\def\art{\@ifnextchar[{\eart}{\oart}}
\def\eart[#1]#2#3#4#5#6{{\rm #2}, {\em #3 \bf #4} {\rm (#6) #5} ({\em #1})}
\def\article{\@ifnextchar[{\earticle}{\oarticle}}
\def\oarticle#1#2#3#4#5#6{{\rm #1}, {\ital ``#6''}, {\rm #2 #3 (#5) #4}}
\def\earticle[#1]#2#3#4#5#6#7{{\rm #2}, {\ital ``#7''}, {\rm #3 #4 (#6) #5}  [\hhrefq{#1}]}
\def\hepart[#1]#2{{\rm #2, \sl#1}}
\def\heparticle[#1]#2#3{#2, {\ital ``#3''} [\hhrefq{#1}]}
\newcommand{\doi}[1]{\href{http://dx.doi.org/#1}{[link]}}

\renewenvironment{thebibliography}[1]
     {\begin{multicols}{2}[\section*{\refname}]%
      \@mkboth{\MakeUppercase\refname}{\MakeUppercase\refname}%
      \list{\@biblabel{\@arabic\c@enumiv}}%
           {\settowidth\labelwidth{\@biblabel{#1}}%
            \leftmargin\labelwidth
            \advance\leftmargin\labelsep
            \@openbib@code
            \usecounter{enumiv}%
            \let\p@enumiv\@empty
            \renewcommand\theenumiv{\@arabic\c@enumiv}}%
      \sloppy
      \clubpenalty4000
      \@clubpenalty \clubpenalty
      \widowpenalty4000%
      \sfcode`\.\@m}
     {\def\@noitemerr
       {\@latex@warning{Empty `thebibliography' environment}}%
      \endlist\end{multicols}}

\newcounter{alphaequation}[equation]
\def\thealphaequation{\theequation\hbox to
0.6em{\hfil\alph{alphaequation}\hfil}}
\def\eqnsystem#1{
\def\@eqnnum{{\rm (\thealphaequation)}}
\def\@@eqncr{\let\@tempa\relax \ifcase\@eqcnt \def\@tempa{& & &} \or
  \def\@tempa{& &}\or \def\@tempa{&}\fi\@tempa
  \if@eqnsw\@eqnnum\refstepcounter{alphaequation}\fi
\global\@eqnswtrue\global\@eqcnt=0\cr}
\refstepcounter{equation} \let\@currentlabel\theequation \def\@tempb{#1}
\ifx\@tempb\empty\else\label{#1}\fi
\refstepcounter{alphaequation}
\let\@currentlabel\thealphaequation
\global\@eqnswtrue\global\@eqcnt=0 \tabskip\@centering\let\\=\@eqncr
$$\halign to \displaywidth\bgroup \@eqnsel\hskip\@centering
$\displaystyle\tabskip\z@{##}$&\global\@eqcnt\@ne
\hskip2\arraycolsep\hfil${##}$\hfil& \global\@eqcnt\tw@\hskip2\arraycolsep
$\displaystyle\tabskip\z@{##}$\hfil
\tabskip\@centering&\llap{##}\tabskip\z@\cr}
\def\endeqnsystem{\@@eqncr\egroup$$\global\@ignoretrue} \makeatother

\definecolor{Gray}{gray}{0.95}

\begin{document}

\begin{center}
{\Large\LARGE \bf \color{rossos}
Direct anthropic bound on the \\[3mm] 
weak scale from supernov\ae{} explosions
}\\[2mm]
{\bf \color{rossos} or: How I Learned to Stop Worrying and Love the Higgs}\\[1cm]

{\bf Guido D'Amico$^{a,b,c}$, Alessandro Strumia$^{d}$, Alfredo Urbano$^{e,f}$, 
Wei Xue$^g$}\\[5mm]

{\it $^a$ Dipartimento di SMFI dell'Universit{\`a} di Parma, Italy}\\[1mm]
{\it $^b$ INFN, Gruppo Collegato di Parma, Italy}\\[1mm]
{\it $^c$ Stanford Institute for Theoretical Physics, USA}\\[1mm]
{\it $^d$ Dipartimento di Fisica dell'Universit{\`a} di Pisa, Italy}\\[1mm]
{\it $^e$ INFN, Sezione di Trieste, SISSA, Italy}\\[1mm]
{\it $^f$ IFPU, Institute  for  Fundamental Physics  of  the  Universe, Trieste,  Italy}\\[1mm]
{\it $^g$ CERN, Theory Division, Geneva, Switzerland}\\[1mm]

\vspace{1cm}

{\large\bf\color{blus} Abstract}
\begin{quote}\large
Core-collapse supernov\ae{}
presumably explode because trapped
neutrinos push the material out of the stellar envelope.
This process is directly controlled by the weak scale $v$:
we argue that supernova explosions happen only if fundamental constants 
are tuned within a factor of few as
$v \sim \Lambda_{\rm QCD}^{3/4} M_{\rm Pl}^{1/4}$,
such that neutrinos are trapped in supernov\ae{}
for a time comparable to the gravitational time-scale.
We provide analytic arguments and simulations in spherical approximation,
that need to be validated by more comprehensive simulations. 
The above result can be important for fundamental physics, because
core-collapse supernova explosions seem anthropically needed, 
as they spread intermediate-mass nuclei presumably necessary for `life'.
We also study stellar burning, finding that it 
does not provide anthropic boundaries on $v$.
\end{quote}

\thispagestyle{empty}
\bigskip

\end{center}
\begin{quote}
{\large\noindent\color{blus} 
}

\end{quote}
\vspace{-1.5cm}

\pagebreak
\tableofcontents

\setcounter{footnote}{0}

\section{Introduction}
Nature contains two relative
mass scales: the vacuum energy density
$V \sim (10^{-30}M_{\rm Pl})^4$
and the weak scale $v^2 \sim (10^{-17}M_{\rm Pl})^2$
where  $v$ is the Higgs vacuum expectation value.
Their smallness with respect to the Planck scale $M_{\rm Pl} = 1.2~10^{19}\GeV$
is not understood and is considered as `unnatural' in relativistic quantum field theory,
because it seems to require precise cancellations among much larger contributions.
If these cancellations happen for no fundamental reason, they are `unlikely',
in the sense that summing random order one numbers gives $10^{-120}$
with a `probability' of about $10^{-120}$.
Worrying about naturalness of the Higgs mass and of the vacuum energy has been
a major theme in fundamental physics in the last decades~\cite{Weinberg:1988cp,hep-th/0603249,1407.2122}.
Many theories alternative to small tuned values have been proposed.
Most theorists expected that the Higgs should have been accompanied by 
new physics that keeps its mass naturally light, 
but experiments discovered just the Higgs~\cite{1407.2122}.
Collider data and cosmological observations 
are so far consistent with small tuned values of the weak scale and of the
cosmological constant.

\smallskip

A controversial but lapalissian anthropic tautology 
seems relevant to understand what goes on in fundamental physics:
observers can only observe physics compatible with their existence.

\smallskip

On the cosmological constant side, its
smallness has been interpreted through an anthropic argument:
a cosmological constant $\sim10^3$ times larger than its physical value would have prevented structure formation~\cite{Wein,SmallLambda}.
No natural theoretical alternatives are known
(for example, supergravity does not select $V=0$ as a special point~\cite{Weinberg:1988cp}), and
anthropic selection of the cosmological constant seems possible
in theories with some tens of scalars 
such that their potential has more than $ 10^{120}$  different vacua, which get
`populated' forming a `multiverse' through eternal inflation.
String theory could realise this scenario~\cite{hep-th/0004134,hep-th/0302219,hep-th/0301240}.


On the Higgs side, it has been noticed that light quark and lepton masses
$m_e$, $m_u$, $m_d$ are  anthropically restricted in a significant way:
a non-trivial nuclear physics with more nuclei than just H and/or He
(and thereby chemistry, and `life') exists because $m_e/\Lambda_{\rm QCD}$, $m_u/\Lambda_{\rm QCD}$, $m_d/\Lambda_{\rm QCD}$  
have appropriate values which allow
for the existence of a hundred of nuclear species~\cite{hep-ph/9707380,0712.2454,hep-ph/0703219,0809.1647}.  
Such anthropic boundaries in $m_e$, $m_u$, $m_d$ give extra indicative support to
the possibility that physics is described by a theory where fundamental constants have different values in different local minima.



\medskip

We point out that, 
however, anthropic selection of fermion masses
$m_f$ does not lead to clean implications for theories of the weak scale.
Rather, it leads to a confusing and paradoxical situation.
Indeed, in the Standard Model,
quark and lepton masses  are given by $m_f = y_f v$ where
$y_f$ are Yukawa couplings, which can be naturally small.
If Yukawa couplings have unique values, 
the weak scale $v$ must be anthropically small:
changing $v$ by a factor of few from its physical value $v_{\rm SM}=174\GeV$ 
changes fermion masses removing complex chemistry~\cite{hep-ph/9707380}.

But the multiple anthropic boundaries on $m_e$, $m_u$, $m_d$
indicate that Yukawa couplings do not have unique values.
Then one looses the anthropic interpretation of  $v \ll M_{\rm Pl}$,
and the Standard Model appears uselessly fine-tuned:
one can easily find anthropically acceptable alternative theories 
less tuned that the Standard Model.
An example  is
a SM-like theory with the same $m_f$, obtained from a bigger $v$ times smaller Yukawa couplings $y_f$.
The paradox is that, in a multiverse landscape,  more tuned vacua are relatively rarer.
We should have expected to live in a less tuned vacuum.  
There are two main classes of less tuned vacua, leading to two aspects of the paradox:
\begin{itemize}
\item[a)] Natural theories: extensions of the SM where $v$ is naturally small compared to $M_{\rm Pl}$.
\item[b)] Less unnatural theories: SM-like theories 
with smaller $y_f$ and bigger $v/M_{\rm Pl}$.
\end{itemize}
\subsubsection*{a) Natural theories}
Collider bounds suggest that nature did not use a natural theory to achieve 
the observed small weak scale.
Unnaturalness of the weak scale, hinted by previous colliders~\cite{hep-ph/9811386,hep-ph/0007265},
has been confirmed by the Large Hadron Collider,
and might be established by future colliders.
We here assume that the weak scale is not natural.

This is not necessarily a paradox: in a generic multiverse context, nature might have avoided a
natural extension of the SM if all such extensions have a
`multiverse probability' so much lower than unnatural models such as the SM,
that the gain in probability due to naturalness 
($\sim 10^{34}$ for the weak scale, possibly times 
$\sim 10^{60}$ for a partial suppression
of the cosmological constant down to the weak scale)
is not enough to statistically favour natural models over unnatural models.
This possibility seems reasonable, given that very few natural extensions of the SM have been proposed, and they employ special ingredients, such as supersymmetry,
which might be rare or absent in a landscape.

\smallskip

In the string context, model building focused on
effective 4-dimensional theories below the string scale where the
weak scale is naturally small thanks to
$N=1$ weak-scale supersymmetry.
If the weak scale is instead unnatural and anthropically selected,
the paradox is avoided provided that 
the landscape distribution of SUSY breaking scales
is dominated by the largest energies
(maybe because breaking supersymmetry dynamically at low energy 
needs somehow contrived model building
in string models with a dilaton),
suggesting that we live in a vacuum with
no supersymmetry below the string scale.

\smallskip

Natural theories based on ingredients different from supersymmetry 
have more evident theoretical problems
that can explain why nature did not use them.
Concerning large extra dimensions, 
the problem is dynamically stabilizing their size in a natural way.
Concerning composite Higgs models,
the only dynamics proposed so far that gives phenomenologically acceptable models
(with Yukawa couplings and thereby fermion masses) 
has similar naturalness issues as the SM itself, as it
employs a fundamental scalar~\cite{1607.01659} with new strong interactions.
Other scenarios where the weak scale is only partially natural involve baroque model-building.





\subsubsection*{b) Less unnatural theories}
The SM appears uselessly more tuned
than similar theories where the anthropic bounds on fermion masses  $m_f = y_f v$
is satisfied using smaller Yukawas $y_f$ and larger $v/M_{\rm Pl}$
(possibly up to the weak-less limit $v \sim M_{\rm Pl}$).
This is a real paradox, because (unlike in the case of natural SM extensions),
it looks not plausible that SM-like theories
have a drastically lower `multiverse probability' than the SM itself.
The SM indicates that some mechanism can generate
small Yukawas such as $y_e \sim 10^{-6}$,
so that it's difficult to argue that smaller Yukawas are highly unlikely.

One way out from this paradox is that the SM itself is natural: this requires
special theories of quantum gravity that do not employ particles much heavier than the Higgs and significantly
coupled to it~\cite{1303.7244,1504.05403,1705.03896}.
A different way out is the possible existence of
an extra anthropic boundary that restricts {\em directly} the scale $v$ of weak interactions.

\subsubsection*{Searching for a direct anthropic boundary on the weak scale}
The Higgs vacuum expectation value $v$
 determines the Fermi coupling
$\GF = 1/(2\sqrt{2} v^2)$ that controls neutrino interactions at low energies,
$\sqrt{s}\circa{<} M_{W,Z}$.
The weak scale $v$ might then
be anthropically relevant in two situations~\cite{Carter:1974zz,Carr:1979sg,Barrow:1988yia} 
where non-trivial physics arises because of a
numerical coincidence.
In both cases the numerical coincidence is 
\beq  \label{eq:vcr1}
u \equiv \frac{M_{\rm Pl}^{1/4} \Lambda_{\rm QCD}^{3/4}}{v}  \sim 1,\eeq
where $\Lambda_{\rm QCD}\approx 300\MeV$ is the QCD scale that
naturally generates nucleon masses $m_{n,p}\sim \Lambda_{\rm QCD}$
through  dimensional transmutation.
In the definition of $u$ we ignored ${\cal O}(1)$ factors, which actually happen to be ${\cal O}(100)$.
The two situations are:
\begin{enumerate}
\item  Big Bang Nucleosynthesis (BBN) predicts an order-one ratio
between the number of neutrons $N_n$ and of protons $N_p$
because the neutrino decoupling temperature $T_{\nu\rm dec}$
is comparable to the the proton-neutron mass difference,
and because the neutron life-time $\tau_n$
is comparable to the age of the Universe at BBN time, $t_{\rm BBN}$~\cite{1409.0551}:
\beq \frac{N_n}{N_p} \approx \exp\left(-\frac{m_n-m_p}{T_{\nu\rm dec}} -\frac{ t_{\rm BBN}}{\tau_n}\right)\approx \frac17.
\eeq
Varying $v$ and $M_{\rm Pl}$ with quark masses fixed,
the first term in the exponent scales as 
$(M_{\rm Pl}/v^4)^{1/3}$ and the second as $M_{\rm Pl}/v^4$:
they depend on the same combination of $M_{\rm Pl}$ and $v$, as in eq.\eq{vcr1}.
A larger (smaller)  $v$  increases (decrease) the He/H ratio,
for not too large variations of $v$,  at fixed $M_{\rm Pl}$ and fixed baryon asymmetry.
Some authors discuss the possibility that a large
$v\circa{>}100v_{\rm SM}$ might be anthropically excluded because
of a too low Hydrogen abundance, as H is used in `life' and molecular H
plays a role in gas cooling that leads to star formation~\cite{astro-ph/0108071,1409.0551}.
However $p$ are not strongly suppressed at BBN;
furthermore extra $p$ can be later produced by stars
(as they make heavy nuclei that
contain more $n$ than $p$ in view of electric repulsion)
as well as by cosmic rays.

\begin{figure}[t]
\begin{center}
\includegraphics[width=0.95\textwidth]{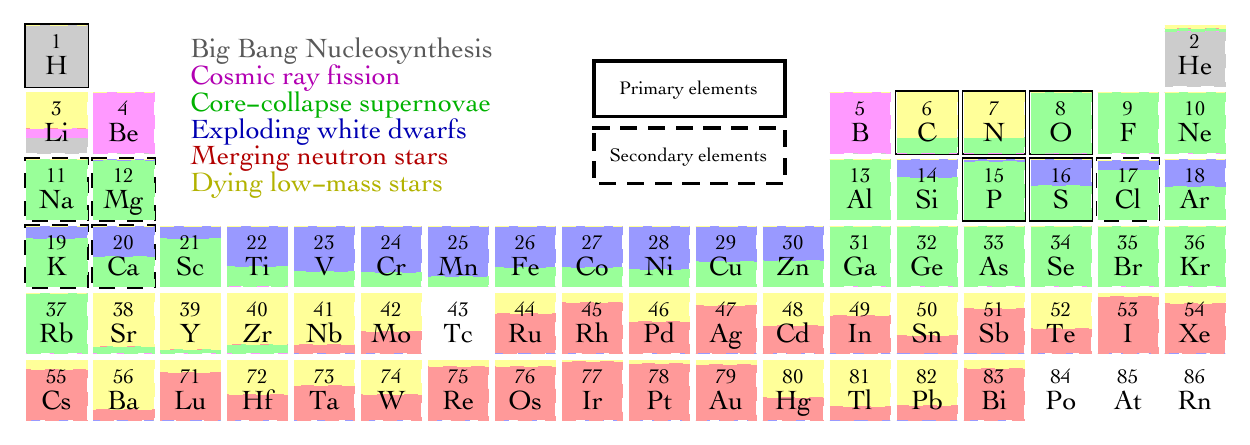}
\caption{\em Each stable element in the periodic table is colored according to its present
relative contributions of nucleosynthesis sources~\cite{Astroper}.
The primary (secondary) elements of terrestrial life are highlighted 
by a continuous (dashed) contour.
\label{fig:PeriodicTable}}
\end{center}
\end{figure}

\item As we will discuss, 
core-collapse supernova explosions crucially depend on $v$, mainly
because neutrinos push the material that surrounds the core, spreading 
intermediate-mass  elements  (O, F, Ne, Na, Mg, Al and possibly N, Cl, K, Ca) which
seem produced almost exclusively in core collapse SN~\cite{Woosley:1995ip,Astroper} and
seem needed for `life'~\cite{Barrow:1988yia}.
\end{enumerate}
Indeed nuclei form as follows:
H and He are dominantly synthesised during BBN; 
core-collapse supernov\ae{} 
(known as type II,  with rarer type Ib and Ic)
lead to the ejection of the shells of burnt star materials, which include relatively light elements.
Cosmic rays (possibly dominantly produced trough supernov\ae explosions) and 
dying light stars (formed thanks to the elements produced by core-collapse
 supernov\ae)
contribute to the production of
relatively light elements (Li$^3$, Be$^4$, B$^5$, C) as well
as to heavy elements thanks to neutron capture.
Merging neutron stars,
explosions of white dwarfs
(accretion supernov\ae{} known as type Ia SN) 
and dying light stars
make elements heavier than Si, in particular Fe~\cite{hep-ph/0609050}.
Accretion supernov\ae{} are binary objects just below the threshold for carbon fusion,
such that capture of extra mass  triggers runaway nuclear reactions, 
heating and giving an explosive melt-down that  proceeds up to the most stable
nucleus, Fe, leaving negligible amounts of intermediate-mass nuclei.\footnote{We do not explore the possibility
that O, rather than Fe, might be the most stable nucleus in vacua with
different values of $m_{u,d}$ or $\alpha_{\rm em}$.}
As a result, nuclei produced almost exclusively by core collapse supernov\ae{} include O
(the primary element of terrestrial life, together with C, N, H, P, S), Na, Mg and possibly K, Ca, Cl
(the secondary elements of terrestrial life).
It has been argued that the chemistry of O is generically needed for `life'~\cite{Barrow:1988yia}.
If true, explosions of core collapse supernov\ae{} are anthropically relevant,
and their existence is related to weak interactions.


%

%
%
Needless to say, neither astro-biology nor the physics of supernov\ae{}  are fully understood and established.
In particular, some authors claim that,
unlike what believed earlier,
neutrinos and weak interactions might be not needed for core-collapse supernov\ae{} explosions,
which can also explode through collapse-induced thermonuclear explosions~\cite{1506.02655,1601.03422,1811.11178}.
If this alternative mechanism is confirmed and if it efficiently spreads nuclei such as O,
our direct anthropic bound on the weak scale would not be present.

\medskip

This paper is structured as follows.
In section~\ref{SN} we discuss how core-collapse SN explosions would behave
for a different value of $v$, keeping fermion masses fixed.\footnote{Light fermion
masses are independently anthropically constrained~\cite{hep-ph/9707380,0712.2454,hep-ph/0703219,0809.1647}.
For $v<v_{\rm SM}$ a top Yukawa coupling larger than one
would be needed to keep the top quark mass fixed;
anyhow, the top quark plays no anthropic role.}
In section~\ref{star} we briefly comment on how star evolution behaves as function of $v$,
finding no anthropic boundaries.
Conclusions are given in section~\ref{concl}.

\section{Supernova explosions}\label{SN}

\subsection{Standard supernova explosions}
We start by summarising the standard theoretical picture of core-collapse supernova explosions.
While not established, it is consistent with the observation of neutrinos from SN1987A 
(see~\cite{hep-ph/0606054,astro-ph/0612072,1206.2503,1702.08713} for reviews).
We keep all fundamental parameters to their physical values and
we provide estimates which exhibit the dependence 
on fundamental parameters $m_n$, $M_{\rm Pl}$, $v$
ignoring order one factors,
in order to later study modified weak scale or Planck mass.
Gravity and weak interactions give competing comparable effects:
due to this coincidence
supernova explosions are a complex phenomenon and
computing order one factors through numerical simulations is needed to understand what happens~\cite{astro-ph/0211194}.
We summarize the results of numerical simulations, and clarify which features 
follow from dynamical adjustments or from
numerical coincidences (as needed to later
consider different values of fundamental parameters).

\subsubsection*{Collapse}

The life of stars proceeds through subsequent stages controlled by a  balance between 
gravity and the energy released in nuclear fusion reactions, that form heavier elements from
lighter ones.
During the first stage, the hydrogen in the core of the star is converted into helium. 
As the hydrogen in the core is exhausted, gravity --- no longer balanced by hydrogen burning --- causes the core to contract.
Hydrogen burning is still active in a shell surrounding the core, made now primarily 
of helium nuclei.
The temperature of the core rises because of the contraction  up to the point where helium fusion begins. 
For stars with low mass $M \circa{<}8\,M_\odot$ the fusion processes end with the creation of an electron degenerate carbon core (they never becomes 
hot enough to ignite carbon fusion), 
that eventually forms a White Dwarf while the outer material drifts off into space forming a planetary nebula,
giving rise to the elements in fig.\fig{PeriodicTable}.
For massive stars with $M \gtrsim 8\,M_{\odot}$, 
 once helium fuel runs out in the core, a further contraction 
 raises the inner temperature sufficiently 
 so that carbon burning begins. 
 Neon, oxygen and silicon burning stages follow similarly.
Each stage is faster than the previous stage because
the temperature is higher: neutrinos carry away more energy,
while nuclear reactions releases less energy.

Eventually, the silicon burning stage causes the formation of an iron core in the innermost   part of the star.
The formation of the iron core stops the fusion chain because iron is the most stable element.
The endpoint of this chain is an onion structure with shells of successively 
 lighter elements burning around an iron core. We denote with $R_{\rm Fe}$ the radius of the iron core. 
 
The iron core of a star is protected against the crushing force of gravity 
 only by the electron degeneracy pressure, and not because of the energy released  by nuclear
  fusion.
 This means that the iron core has a
maximum mass determined by the Chandrasekhar limit, $M_{\rm Ch} \approx 1.4 M_{\odot}$.
As the silicon shell surrounding the iron core continues to burn, 
the iron core mass
slowly increases approaching the Chandrasekhar limit above which the 
the iron core starts collapsing under its weight.
In terms of fundamental parameters the Chandrasekhar mass is given by
\beq M_{\rm Ch}\sim Y_e^{2}M_{\rm Pl}^3/m_n^2\eeq
where $Y_e\sim 1/2$ is the electron fraction per nucleon.


\subsubsection*{Deleptonization}
\begin{figure}[t]
\begin{center}
\includegraphics[width=0.5\textwidth]{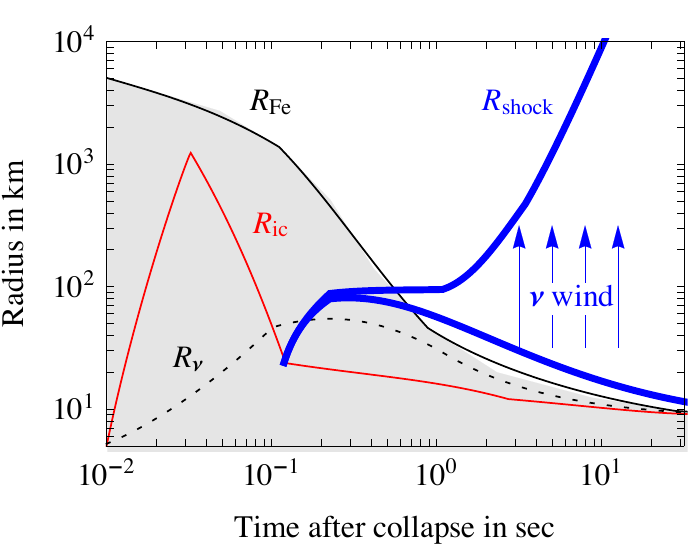}
\caption[Supernova explosion]{\em Sketchy plot of the time evolution of the
main characteristic radii: 1) $R_{\rm Fe}$ is the radius of the iron core;
2) $R_{\rm ic}$ is the radius of the inner iron core which starts collapsing;
3) $R_\nu$ is the radius of the `neutrino-sphere';
4) $R_{\rm shock}$ is the radius of the bounce shock, that stalls and gets 
possibly revitalised through neutrinos emitted from the neutrino-sphere.
\label{fig:SNcollapse}}
\end{center}
\end{figure}

Weak interactions contribute to star burning, and
start playing a crucial role when the 
inert iron core with radius $R_{\rm Fe}$ of a few thousand of km
and mass $M\sim M_{\rm Ch}$ reaches
a high enough sub-nuclear density, $\rho \sim 10^{-6} m_n^4$,
that electrons and protons are converted into neutrons and neutrinos~\cite{Colgate:1966ax}.
Their main  inverse $\beta$ decay reaction is $e+{} ^{56}{\rm Fe}\to {}^{56}{\rm Mn}+\nu_e$,
kinematically allowed at energies  larger than $M_{\rm Mn}- M_{\rm Fe}=3.7\MeV$.
While electrons support the core with their degeneracy pressure,
neutrinos  freely escape.
The core is de-leptonized in a time scale
$\tau_{\rm weak} \sim 1/\sigma_{\rm weak} n_n \propto v^4$ faster than
the free-falling time-scale of the gravitational  collapse
\beq \tau_{\rm grav} \sim\frac{ M_{\rm Pl}}{\sqrt{ \rho}} \sim \frac{M_{\rm Pl}}{m_n^2}
\qquad\hbox{for $\rho\sim m_n^4$}.\eeq
Detailed computations find $\tau_{\rm weak}\approx 10^{-3}\,{\rm sec} \ll \tau_{\rm grav} \approx  0.1\,{\rm sec}$.
As a consequence $Y_e$  and thereby $M_{\rm Ch}$ suddenly decrease:
after an order one decrease, a order one inner fraction of the iron core (called `inner core') starts collapsing.
More precisely, the iron core breaks into two distinct regions: a subsonically and homologously 
collapsing inner core, with mass 
$M_{\rm ic}$ and radius $R_{\rm ic}$, and a supersonic outer core.
In the inner core the inward collapse velocity is proportional to the radial distance 
(giving a  `homologous' collapse), and its boundary $R_{\rm ic}$ is defined as the point where
 the inward radial velocity equals the sound speed of the fluid (which, on the contrary, decreases with density and, 
 therefore, radial distance).
The time evolution of the inner core radius $R_{\rm ic}$ is sketched as a red curve in fig.\fig{SNcollapse}.

\subsubsection*{Rebounce}
When the collapsing inner core (with mass $M_{\rm ic} \approx Y_e^2 M_{\rm Pl}^3/m_n^3$,
more precisely equal to about $0.6M_\odot$ according 
to simulations)
reaches nuclear density the collapse is halted by 
the nuclear force, which is repulsive at nucleon distances below 1 fm.
At this point the radius of the inner core collapsed down to about 
$R_{\rm ic}\approx  20\, \mbox{ km}$.
This can be estimated as $R_{\rm ic} \sim  \hbox{fm}\ N^{1/3} \sim M_{\rm Pl}/m_n^2$
taking into account that there are $N \sim M_{\rm Ch}/m_n \sim (M_{\rm Pl}/m_n)^3 \sim 10^{57}$ nucleons
at distance ${\rm fm}\sim 1/m_n$.
The inner core radius is parametrically the same as the Schwarzschild radius $R_{\rm Sch} = 2M_{\rm Ch}/M_{\rm Pl}^2$.
Simulations find $R_{\rm Sch} > R_{\rm ic}$ for $M \circa{<} 40 M_\odot$, such that
only heavier neutron stars collapse directly into black holes.

If the collapse does not proceed with the formation of a black hole, 
the inner core, being in sonic communication throughout the process, coherently bounces~\cite{Colgate:1966ax}.
The rebound of the inner core generates 
an outward-going shock wave (blue curve in fig.\fig{SNcollapse}) propagating through the still 
infalling outer core.
To generate a supernova explosion, the shock wave must cross the outer iron core.
In doing so, the  shock wave loses energy
due to iron dissociation (giving $8.8~\mathrm{MeV}$ binding energy per nucleon) and 
neutrino production via electron capture on the way through the outer core:
the shock slows down and stalls at $R_{\rm shock}\approx 100\,{\rm km}$.
Order one factors, such as the distinction between the inner and outer core,
are important for causing the stall.
After crossing the iron core with radius $R_{\rm Fe}$ the temperature is not high enough to dissociate nuclei, 
and the shock stalls. 

\smallskip

Weak interactions give a second crucial effect:
the stalled shock wave is rejuvenated by the outflowing neutrinos.

\subsubsection*{Neutrino trapping}

As matter density increases, neutrinos get momentarily trapped in the collapsing star,
up to a neutrino-sphere radius $R_\nu$ which is slightly larger than the inner core radius $R_{\rm ic}$
(dashed and red curves, respectively, in fig.\fig{SNcollapse}),
because of a numerical coincidence which involves the Fermi and Newton constants.
Thereby the gravitational energy produced by the collapse, $E_{\rm tot} \sim \GN M^2_{\rm ic} /R_{\rm ic} \sim M_{\rm Pl}^3/m_n^2 \sim 3~10^{53}\,{\rm erg} $,
remains trapped behind the shock wave, and is released as neutrinos
with a time-scale (comparable to the  time-scale of the shock, about a second),
given by
\beq \label{eq:taucool}\tau_{\nu} \sim \max(\tau_{\rm volume}, \tau_{\rm surface})\eeq
where the two factors are the time-scales for energy transmission
inside the trapping volume, and from its neutrino-sphere surface.

The first factor is controlled by diffusion of trapped neutrinos:
the neutrino cross section is $\sigma_{\rm weak} \sim T^2/v^4$ such that the neutrino
mean free path at temperature $T_{\rm ic} \sim m_n$ is
\beq \ell_\nu \sim \frac{1}{n_n\sigma_{\rm weak}} \sim \frac{v^4}{m_n^3T_{\rm ic}^2} .\eeq
The diffusion time of neutrinos inside the 
inner core can be computed in a simple way in terms of random walk, 
given that the inner core has a constant matter density.
A neutrino covers a distance $R_{\rm ic}$ in $N_\nu \sim  R_{\rm ic}^2/\ell^2_\nu$ diffusion steps.
Neutrinos diffuse on a time 
\beq \tau_{\rm volume} \sim \max(N_\nu,1) \ell_\nu \sim
\max\bigg(\frac{m_n M_{\rm Pl}^2}{v^4} ,\frac{M_{\rm Pl}}{m_n^2} \bigg).\eeq

\smallskip

The second factor in eq.\eq{taucool}, $\tau_{\rm surface}$, 
depends on the radius $R_\nu $ of the neutrino-sphere:
as it is bigger than the inner core radius $R_{\rm ic}$,
we need to know the profile density of the supernova outside the inner core.
Simulations find that the nucleon number density $n(r)$ varies outside the inner core (after that it is stabilised) as
\beq \label{eq:nR3}
n \sim n_{\rm ic} (R_{\rm ic}/r)^3,\qquad n_{\rm ic} \sim m_n^3.\eeq
The temperature of the material is
given by the Fermi momentum, $T  \sim n^{1/3}$ and thereby
scales as  $T \sim T_{\rm ic} R_{\rm ic}/r$, having assumed
that conduction of energy inside the inner core controlled by $\tau_{\rm volume}$
is fast enough to compensate energy losses from the surface
(otherwise the surface temperature gets lower).
Imposing $n_n(R_\nu) \sigma_{\rm weak} R_\nu \sim 1$ determines the radius of the
neutrino-sphere
\beq\label{eq:RnuGF}
R_\nu \sim \frac{n_{\rm ic}^{1/4} R_{\rm ic}^{5/4} T_{\rm ic}^{1/2}}{v} 
\sim
 \frac{1}{v} \left(\frac{M_{\rm Pl}}{m_n}\right)^{5/4},\qquad
T_\nu \sim  v \left(\frac{m_n}{M_{\rm Pl}}\right)^{1/4}.\eeq
The power emitted in neutrinos thereby is 
\beq \label{eq:Lnu}
L_\nu \sim R_\nu^2 T_\nu^4 \sim v^2 (M_{\rm Pl}/m_n)^{3/2},\eeq 
and the
cooling time of the surface is
\beq \label{eq:taucoolGF}
\tau_{\rm surface}  = \frac{E_{\rm tot}}{L_\nu} \sim \frac{M_{\rm Pl}^{3/2}}{v^2m_n^{1/2}}.\eeq
Detailed numerical computations including order one factors 
(such as $n_{\rm ic} \approx (0.2 m_n)^3$, $T_{\rm ic}\approx 0.1 m_n$, \ldots)
find  
$\ell_\nu \sim \hbox{few cm} < R_{\rm ic}$ so that neutrinos are trapped.
The various time-scales happen to be comparable with the fine structure 
$\tau_{\rm volume}  \sim {\rm sec} \circa{>} \tau_{\rm grav},\tau_{\rm surface}$
for the time-scale over which neutrinos emitted from the neutrino-sphere drain the gravitational energy of the collapse.

\subsubsection*{Heating due to out-going neutrinos}

Nucleons immediately outside the neutrino-sphere
are heated with rate~\cite{Bethe:1984ux}
\begin{equation}\label{eq:Heat1}
Q_{\nu} \sim \sigma_{\rm weak}  \frac{ L_{\nu}}{4\pi R_\nu^2}\sim v^2\left(
\frac{M_{\rm Pl}}{m_n}
\right)^{3/2}\qquad\hbox{where}\qquad \sigma_{\rm weak} \sim \GF^2 T_{\nu}^2.
\end{equation}
More precisely  the interactions of neutrinos with matter 
outside the neutrino-sphere
result in two antagonistic processes: cooling and heating. 
On the one hand, outgoing neutrinos heat
free nucleons (present at $r<R_{\rm shock}$)
more efficiently than nuclei (present at $r>R_{\rm shock}$)
through weak reactions
 $n+\nu_e\to e^- + p$ and $p+\bar{\nu}_e \to e^+ + n$.
The heating rate per nucleon $Q^+_{\nu}$ at generic
radial distance $R_{\nu}\leqslant r \leqslant R_{\rm shock}$  is estimated as
\begin{equation}\label{eq:Heat}
Q^+_{\nu}(r) \sim \frac{\GF^2 L_{\nu}\langle E_{\nu}^2\rangle}{4\pi r^2}\sim \frac{1}{r^2}
\frac{M_{\rm Pl}}{m_n},
\end{equation}
where $\langle E_{\nu}^2\rangle\sim T_\nu^2$ is the mean squared energy of neutrinos.  
On the other hand, nucleons cool down by radiating neutrinos as a consequence of 
electron and positron capture with a typical 
cooling rate 
\begin{equation}\label{eq:Cool}
Q^-_{\nu}(r) \sim \GF^2 T^6 ,
\end{equation}
where $T \sim T_{\rm ic} R_{\rm ic}/r$ is the temperature of the material,
as already discussed above.
The net heating rate per nucleon due to neutrinos is given by $Q_{\nu}\equiv Q^+_{\nu}-Q^-_{\nu}$.
Cooling typically dominates at small radial distances where the material is hotter but, 
since $Q^+_{\nu}$ decreases less steeply with $r$ than $Q^-_{\nu}$, 
neutrino heating dominates over energy losses above
some gain radius $r>R_{\rm gain}$ given by
\begin{equation}
R_{\rm gain} \sim \frac{1}{v}\left(
\frac{M_{\rm Pl}}{m_n}
\right)^{5/4}.
\end{equation}
Having omitted order one factors we find 
$R_{\rm gain}\sim R_\nu$: in this approximation the expression
for the gain radius is more simply found imposing $T_\nu \sim  T$
irrespectively of the specific processes that dominate energy exchanges.
Heating and cooling would be in equilibrium
if matter at radius $r$ had temperature
$T_{\rm eq} = T_\nu\sqrt{R_{\rm gain}/r}$. 
This is hotter than $T = T_\nu R_{\rm gain}/r$ at $r>R_{\rm gain}$, confirming that
neutrinos heat matter at $r>R_{\rm gain}$ and cool matter at
$r<R_{\rm gain}$.

The presence of the region with positive net heating rate is considered crucial for a successful revival of the 
stagnant shock.
Numerical computations find $R_{\rm gain}\approx 3 R_\nu$ such that
$R_\nu < R_{\rm gain} < R_{\rm shock},R_{\rm Fe}$: outgoing
neutrinos can push the stalled shock wave from below in a way considered 
crucial for finally getting a `delayed explosion'. 
If at least a few \% of the gravitational energy $E_{\rm tot}$
emitted in neutrinos is transferred to the shock 
wave, it explodes the whole star, spreading its nuclei.
This fails by a small margin ($10-20\%$)
according to simulations done in spherical approximation~\cite{astro-ph/0006418}.
Successful explosions seem to require taking into account asphericity and possibly
rotation,  magnetic fields, etc~\cite{astro-ph/0312633,astro-ph/0507135,1212.1747,1409.5779,1505.05110,1506.05139,1807.07579,1809.05106}.
The conclusion seems that
stars with mass $8 M_\odot  \circa{<} M \circa{<} 40 M_\odot$ 
can make SN explosions thanks to neutrinos.\footnote{First stars with low metallicity and large mass $130< M/M_\odot \circa{<} 250$ are believed to undergo explosions 
when photons get energetic enough to produce $e^+ e^-$ pairs, removing pressure support,
such that the consequent compression triggers  runaway nuclear fusion.
}
It is believed that most core-collapse SN explode,
that $99\%$ of their energy is emitted in neutrinos and $1\%$ in other particles, 
spreading elements needed for life.


\subsection{Supernova explosions for different $v$ and $M_{\rm Pl}$: analytic discussion}\label{analytic}
In the previous discussion we provided simple analytic expressions in terms of fundamental parameters,
$v$, $M_{\rm Pl}$, $m_n$.
These analytic expressions help understanding what happens if such fundamental parameters had
values different from their physical values.

Neutrino-induced delayed supernova explosions can only arise if neutrinos are trapped, such that the 
gravitational energy of the inner core is released gradually pushing the shock wave.
This condition can be written either as $R_\nu \circa{>} R_{\rm ic}$
or as  $\tau_{\nu}\circa{>}\tau_{\rm grav}$  or as $\tau_{\rm volume}\circa{>}\tau_{\rm grav}$
or as $N_\nu \circa{>} 1$.  
These conditions give the critical value of eq.\eq{vcr1}:
neutrinos are trapped if
\beq v < v_{\rm trap} \equiv {\cal O}(1)\times m_n (M_{\rm Pl}/m_n)^{1/4}.\eeq
The ${\cal O}(1)$ coefficient 
happens to be $\sim 0.01$ such that $v_{\rm trap}$ is a factor of few
above $v_{\rm SM}$.

\subsubsection*{Supernova explosions for smaller/larger Fermi constant}

Changing $v$ affects the initial deleptonisation phase, and, more importantly, 
the final phase that leads to the explosion.

Concerning deleptonisation, it takes place at the physical value of $v$ because
weak interactions are faster than gravity, $\tau_{\rm weak} \sim0.01 \tau_{\rm grav}$.
This remains true until $v\circa{<} 3 v_{\rm SM} $: we expect that for such values of $v$
the decrease of $Y_e$ keeps happening so fast that only the inner core collapses, 
such that the shock wave in all cases needs to cross roughly the same amount of outer material.
We expect that deleptonisation also happens for larger $v\circa{>} 3 v_{\rm SM} $ 
because the gravitational collapse raises the temperature and density 
increasing the weak interaction rate until deleptonisation of the core finally happens.
Increasing $v$ even more,
massive enough stars reach temperatures and densities where the relevant physics
is QCD rather than nuclear physics: we then expect that there is no rebounce.
Finally, no deleptonisation can occur if $v$ is so large that
the extreme weak-less limit ($\GF=0$) is relevant.
This was studied in~\cite{hep-ph/0604027,hep-ph/0609050}, that  suggested that core-collpase
supernova explosions do not occur.\footnote{The authors of~\cite{hep-ph/0604027} suggested that core-collpase supernov\ae{}  can be replaced by accretion supernov\ae{}, which however do not spread some of the light elements which seem needed for `life'.}
This would need to be established through numerical simulations.\footnote{An alternative
possibility is that the inner core still rebounces and its mass
$M_{\rm ic}\sim Y_e^{2}M_{\rm Pl}^3/m_n^2$ is
bigger than in the physical case (because deleptonizaton does not reduce $Y_e$):
this might allow the shock wave to avoid stalling 
even without the help of outflowing neutrinos.}

\medskip

Coming now to the final neutrino-driven core-collapse supernova explosion
we expect that it is more critically affected by $v$ and that explosions  arise in a narrow range of $v$
\beq v_{\rm min} < v < v_{\rm max} < v_{\rm trap}\eeq
The reason is the following.

\begin{enumerate}

\item If $v > v_{\rm trap}$ neutrinos are not trapped so they escape immediately
having negligible interactions with matter outside:
neutrinos cannot rejuvenate the shock that would lead to a supernova explosion.

\item 
If $ v < v_{\rm trap}$ neutrinos are trapped, and one needs to study
if neutrinos can trigger a supernova
explosion.
In first approximation, the
{\rm total} energy and momentum transmitted by outward-going neutrinos to matter outside
does not depend on $v$ and 
is of order of (a few percent of) the total energy $E_{\rm tot}$.
The reason is that, for any $v$ and for any matter density profile, 
neutrinos undergo about a
scattering after exiting the neutrino-sphere.
On the contrary, the spatial and temporal structures
of neutrino heating depend significantly on $v$:
both $R_{\rm gain} \sim \hbox{few}\times R_\nu\propto1/v$ (see eq.\eq{RnuGF})
and $\tau_{\nu}$  (see eq.\eq{taucool})
grow with $\GF$.

%


%
%

\item Supernovae explosions take place in the usual way
for  $v_{\rm min} \circa{<} v\circa{<} v_{\rm max}$:
the collapsing inner core gets halted by nuclear repulsion, giving a rebounce shock wave that stalls and
is rejuvenated by neutrinos.
This process is maximally efficient
when the time-scale of neutrino cooling is comparable to the
time-scale of the shock.
In such a case, the shock 
can reach the gain region, 
$R_{\rm shock}\circa{>}R_{\rm gain}$.
This happens around the physical value of $v$.
For larger $v_{\rm max} < v<v_{\rm trap}$ neutrinos escape too
fast and deposit energy to more interior regions subject to larger gravitational potentials:
on the time-scale relevant for the shock
the net effect of neutrinos is cooling the shock.

\item We expect no supernova explosion in the opposite limit where
$v$ becomes too small, $ v\circa{<} v_{\rm min}$,
because neutrinos interact so much that
neutrino energy is released on a time-scale much longer than the time-scale
of the shock-wave.
Furthermore, at even smaller $v$
the gain radius and/or the neutrino-sphere radius $R_\nu$ become 
bigger than the radius reached by the stalled shock wave, 
so that shock is cooled and/or not pushed. 
If supernova explosions get prevented by $R_{\rm shock} < R_{\rm gain}$,
we expect that this roughly happens at
$ v_{\rm min}\sim 0.2 v_{\rm SM}$.

\end{enumerate}

Numerical simulations indicate that explosions happen at the physical value of $v$ only
if asphericity is taken into account.
Simulations in spherical approximation give explosions only if 
the neutrino luminosity is artificially enhanced by $10-20\%$~\cite{1010.5550,1501.02845}.
Possibly $v_{\rm max}$ could be ${\cal O}(20\%)$ bigger than the physical Higgs vev $v_{\rm SM}$, but computing its value better than an order-of-magnitude estimate
would need dedicated simulations.

\begin{figure}[t]
$$
\includegraphics[width=0.48\textwidth]{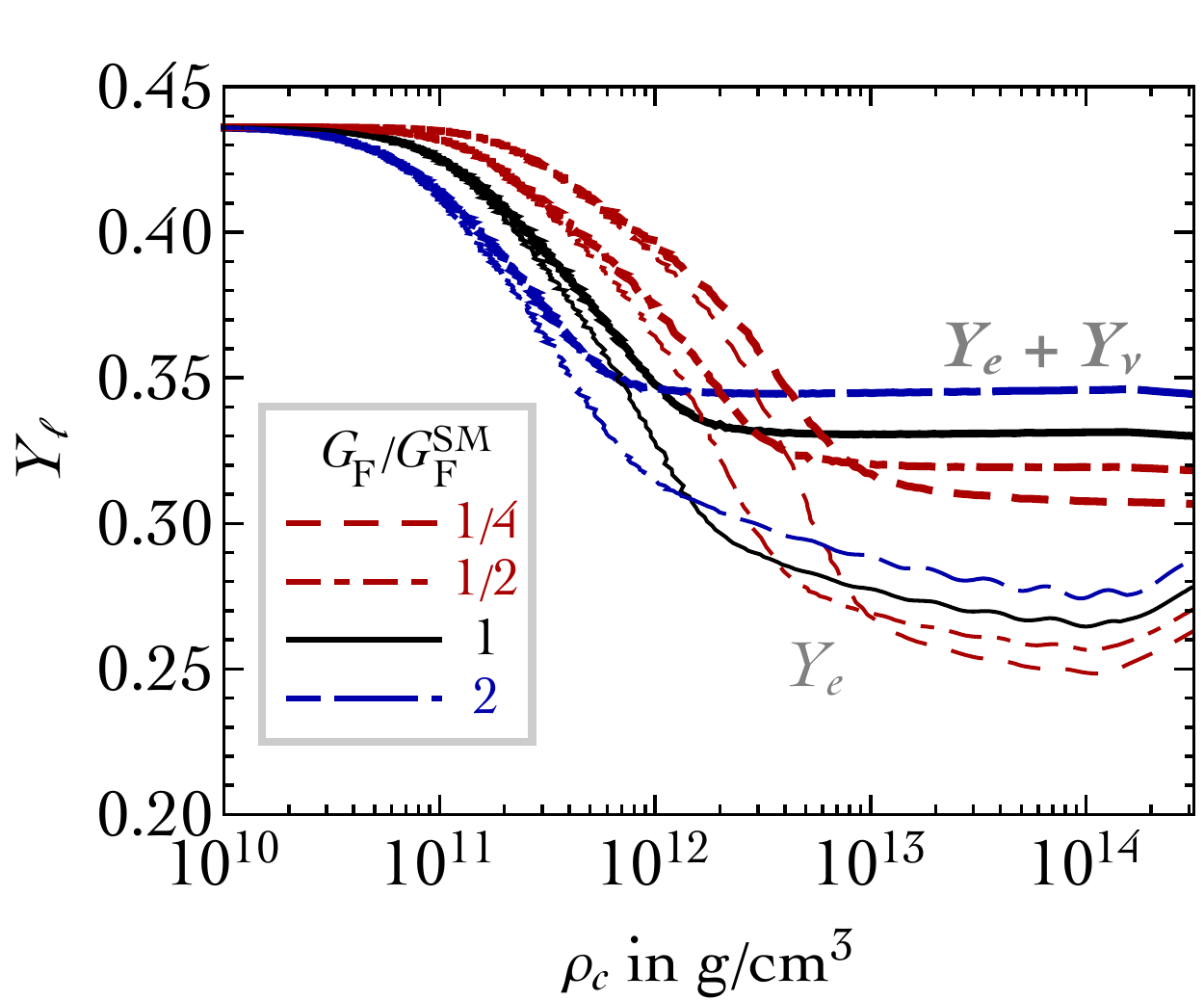}\qquad
\includegraphics[width=0.45\textwidth]{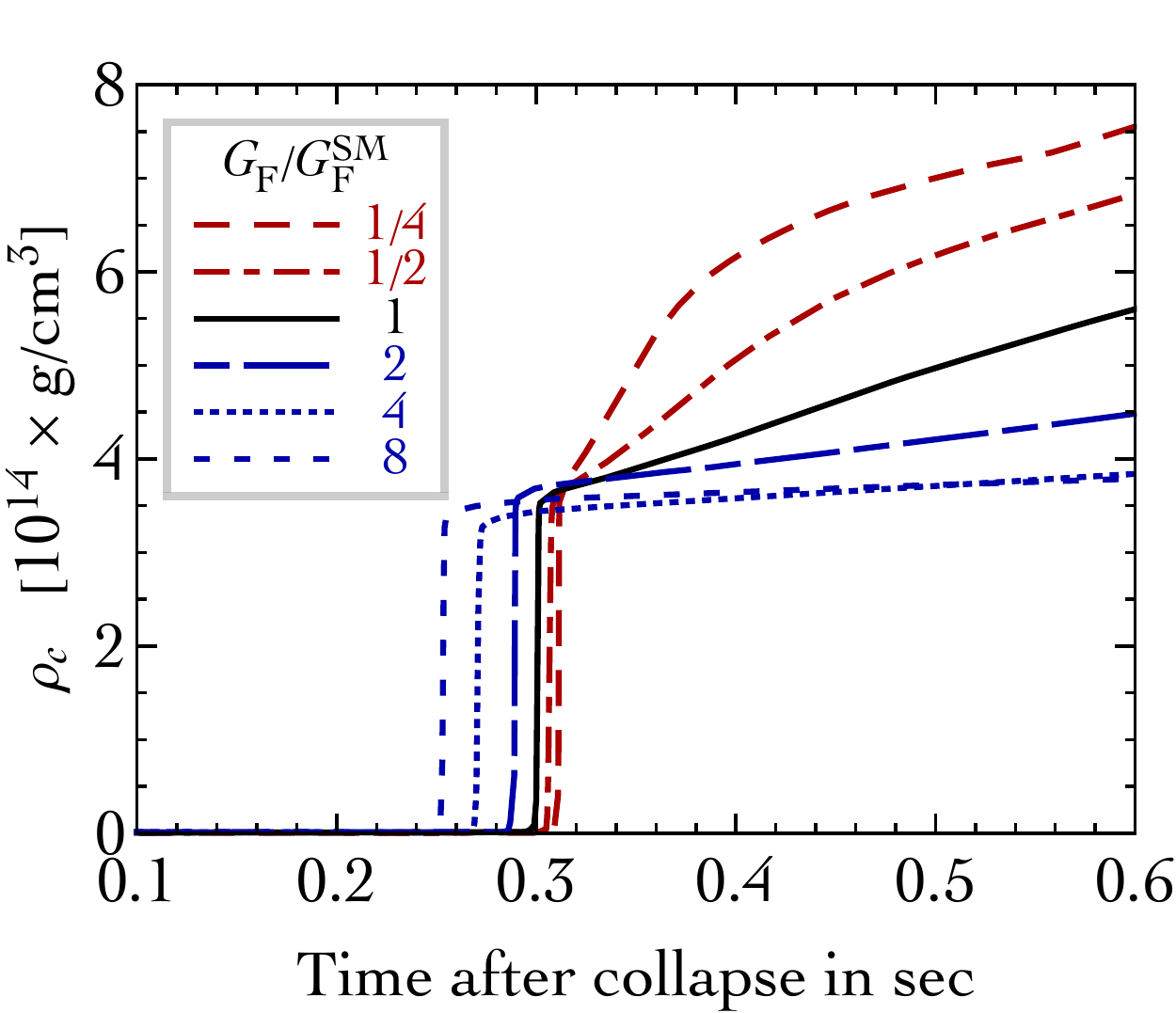}$$
\caption{\em 
{\bf Left:} electron fraction $Y_e$ (thin curves) and total lepton fraction $Y_\ell=Y_e + Y_\nu$
(thick curves)
as function of matter density in the SN center, in the pre-bounce phase after collapse.
{\bf Right:}
Time evolution of the density in the SN center 
for different values of the Fermi constant
$\GF=\{{\color{red}1/4,1/2},1,{\color{blue}2,4,8}\} G_{\rm F}^{\rm SM}$.
We simulated a SN with $M=15M_\odot$ running the code in~\cite{0912.2393}.
\label{fig:Ye}}
\end{figure}

\begin{figure}[!htb!]
\minipage{0.33\textwidth}
\centering
  \includegraphics[width=.95\linewidth]{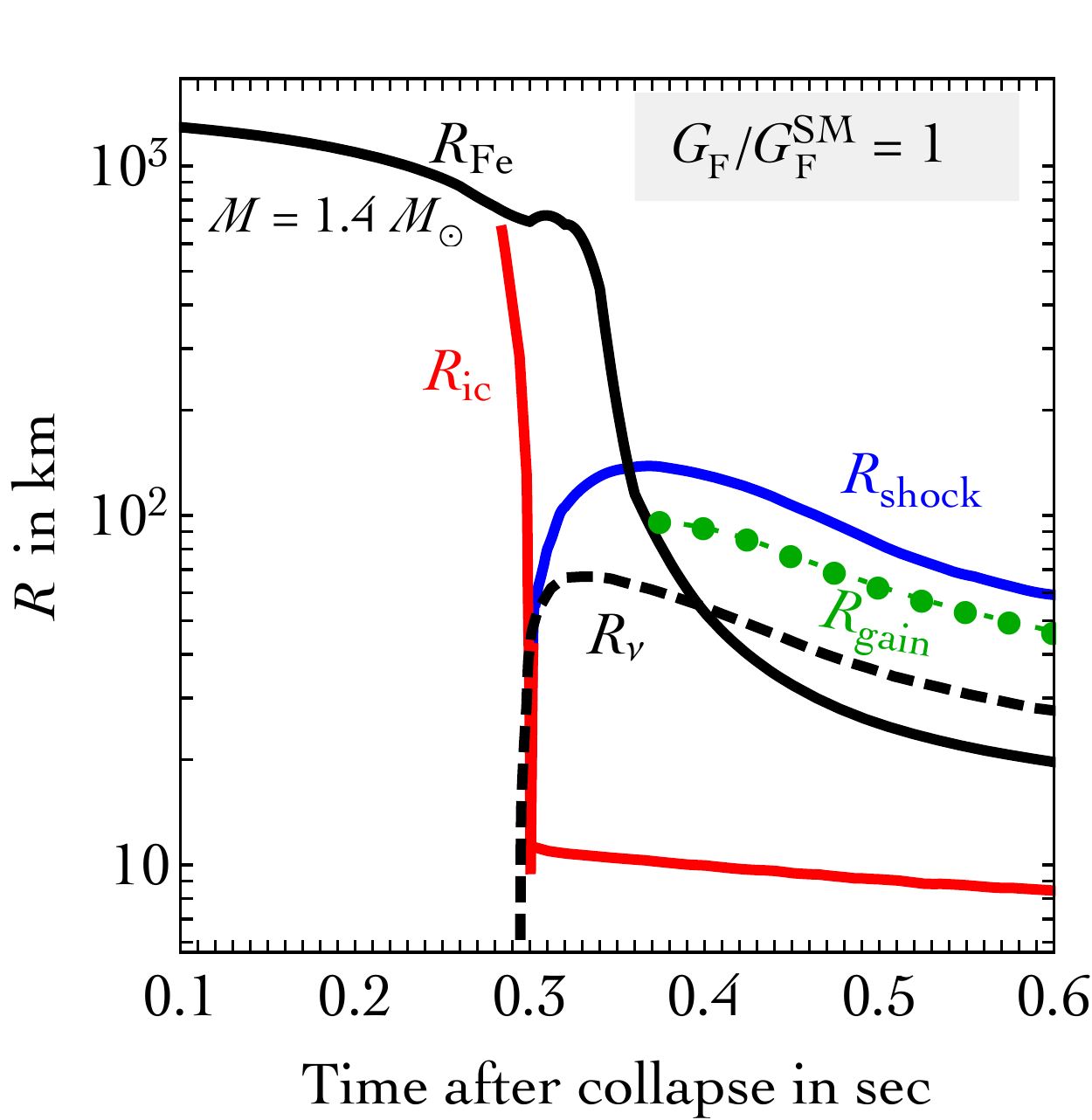}
\endminipage 
\minipage{0.33\textwidth}
  \includegraphics[width=.95\linewidth]{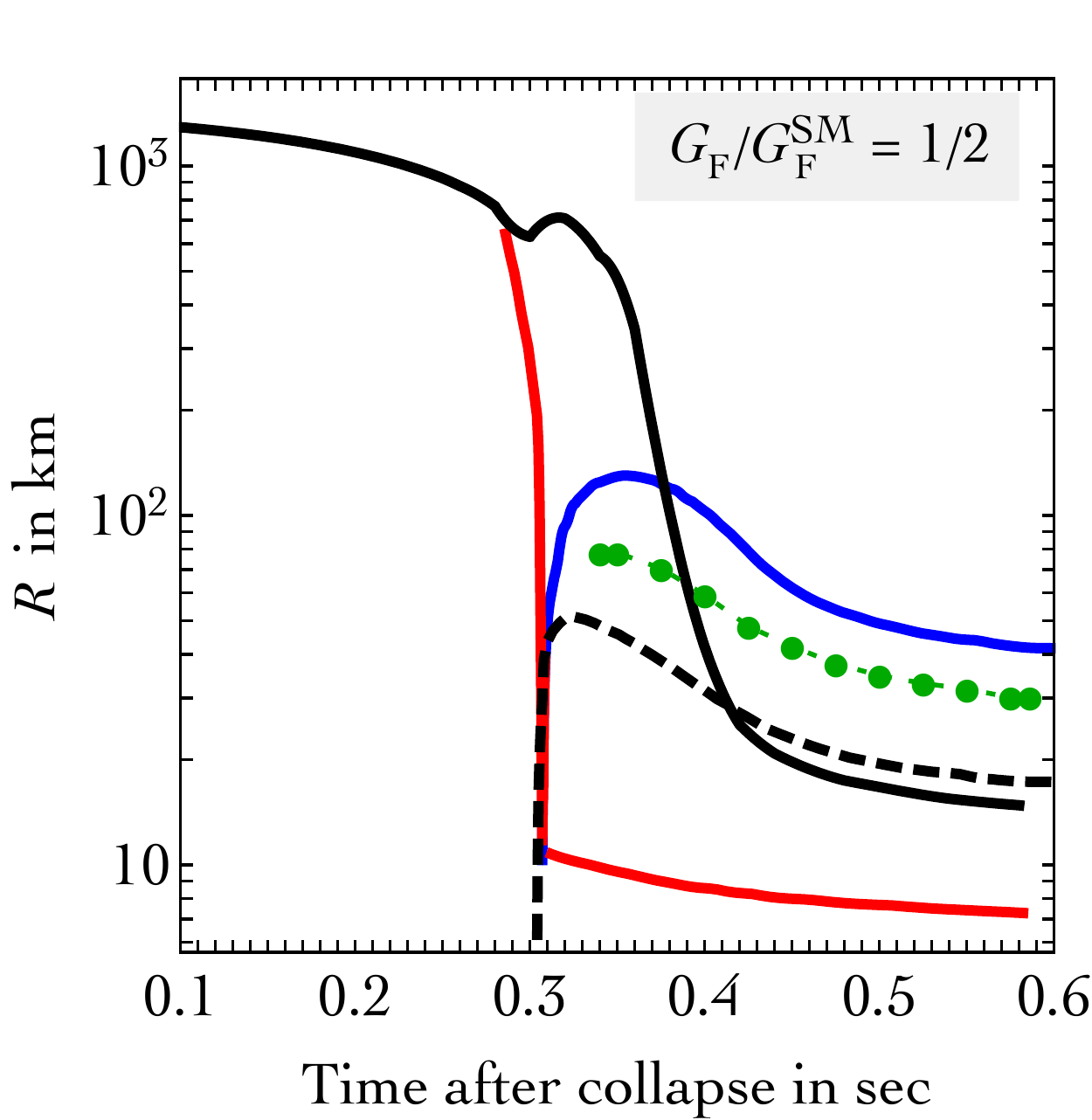}
\endminipage 
\minipage{0.33\textwidth}
  \includegraphics[width=.95\linewidth]{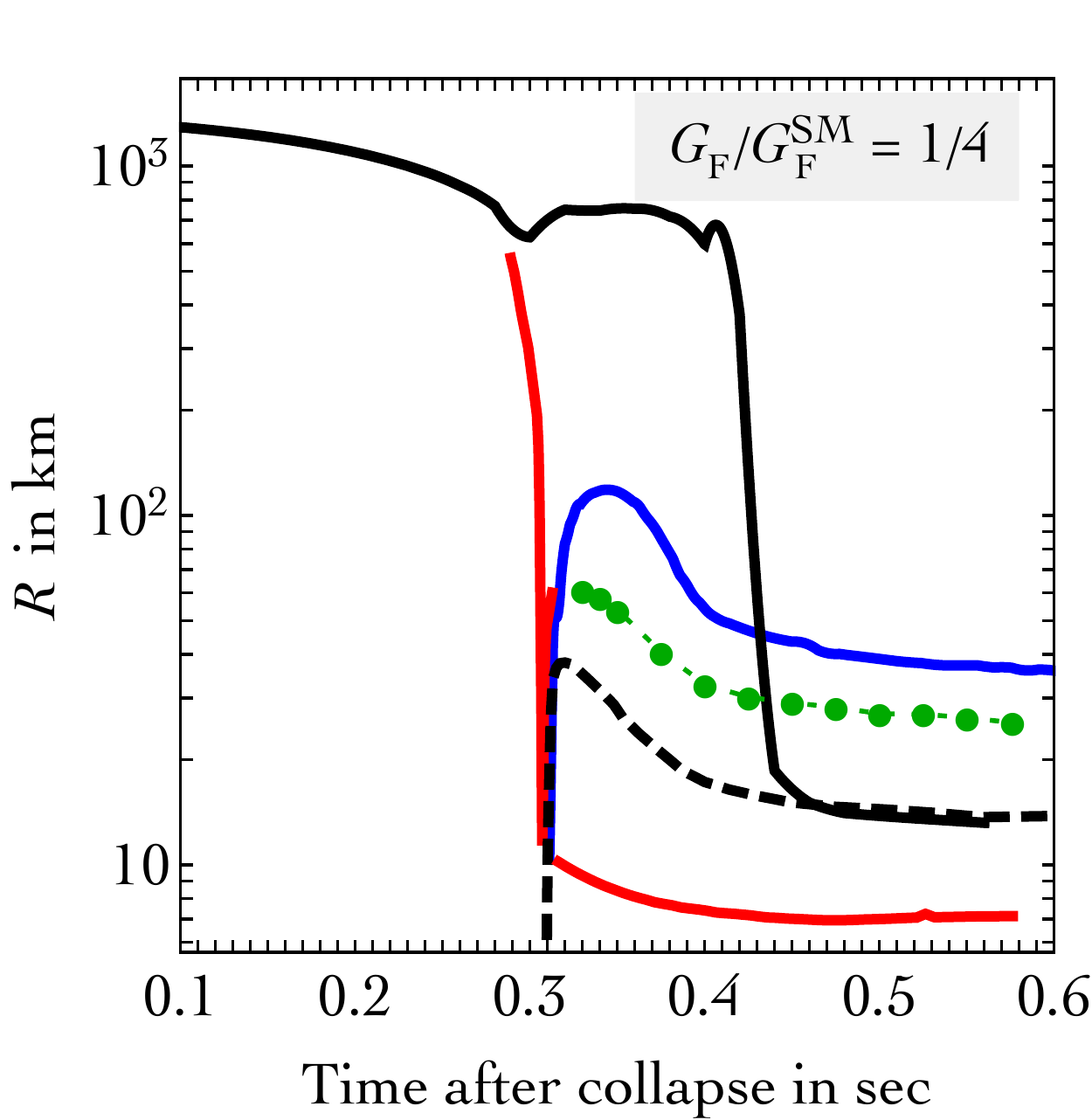}
\endminipage \\
\minipage{0.33\textwidth}
\centering
  \includegraphics[width=.95\linewidth]{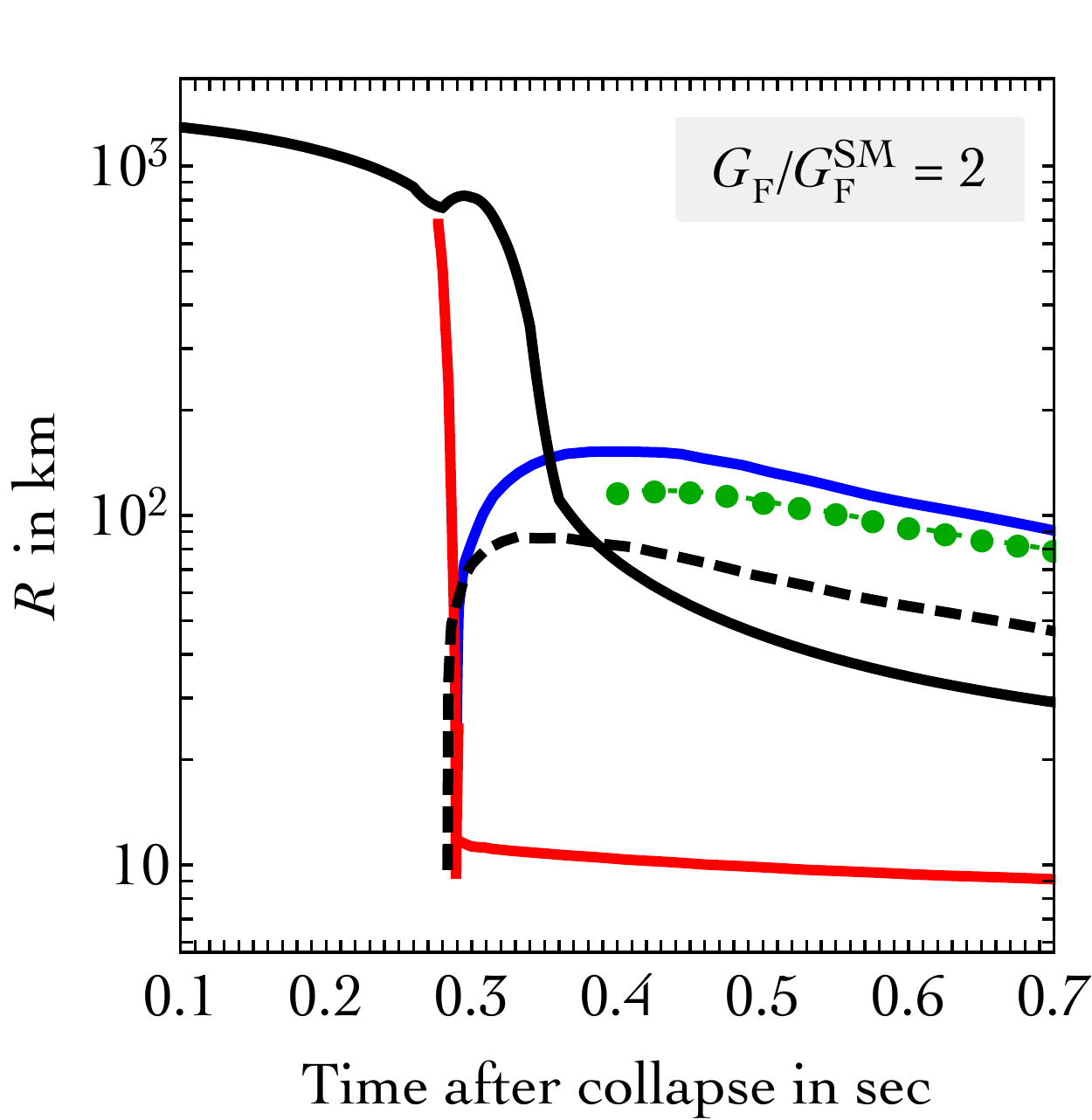}
\endminipage 
\minipage{0.33\textwidth}
  \includegraphics[width=.95\linewidth]{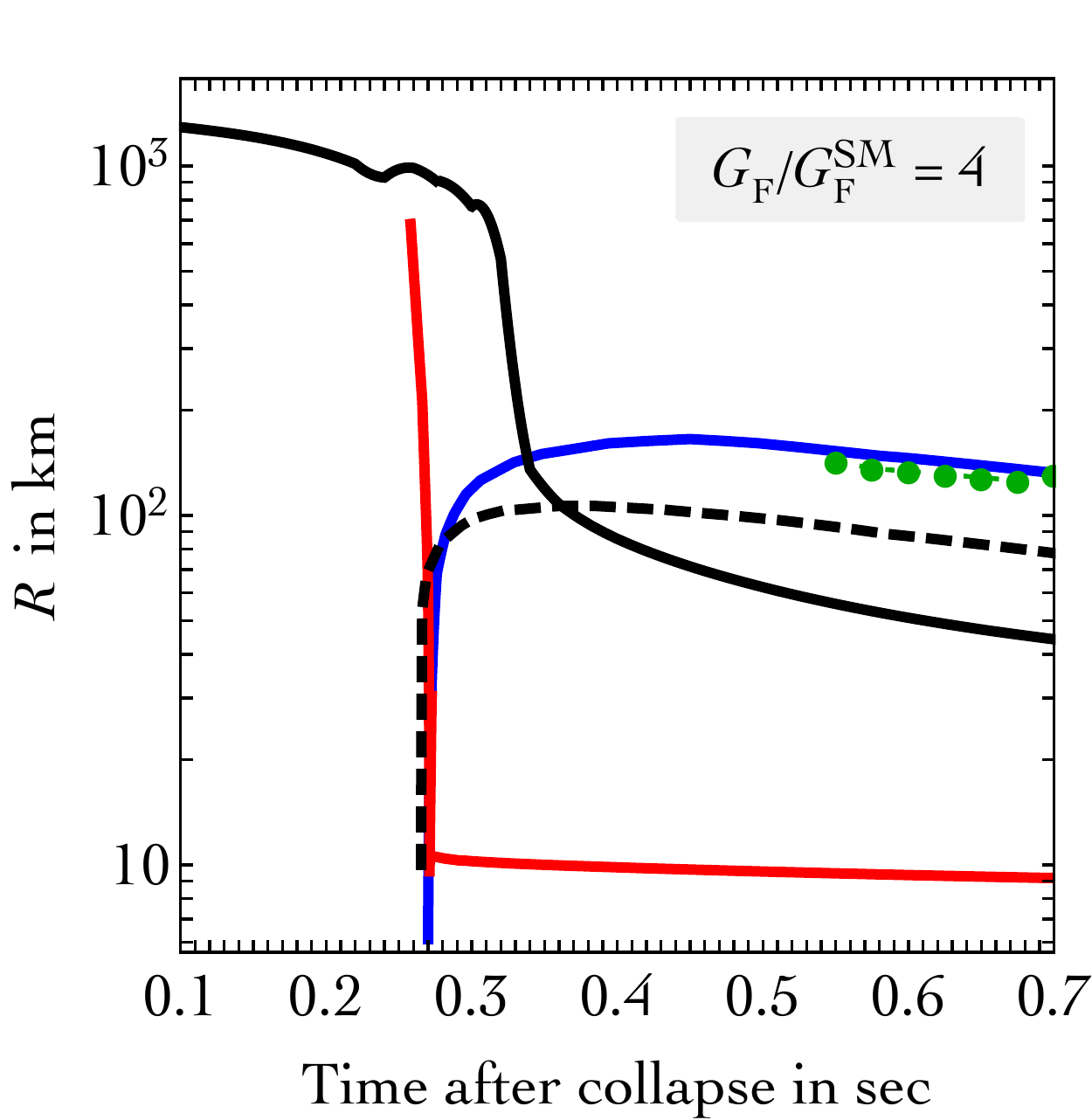}
\endminipage 
\minipage{0.33\textwidth}
  \includegraphics[width=.95\linewidth]{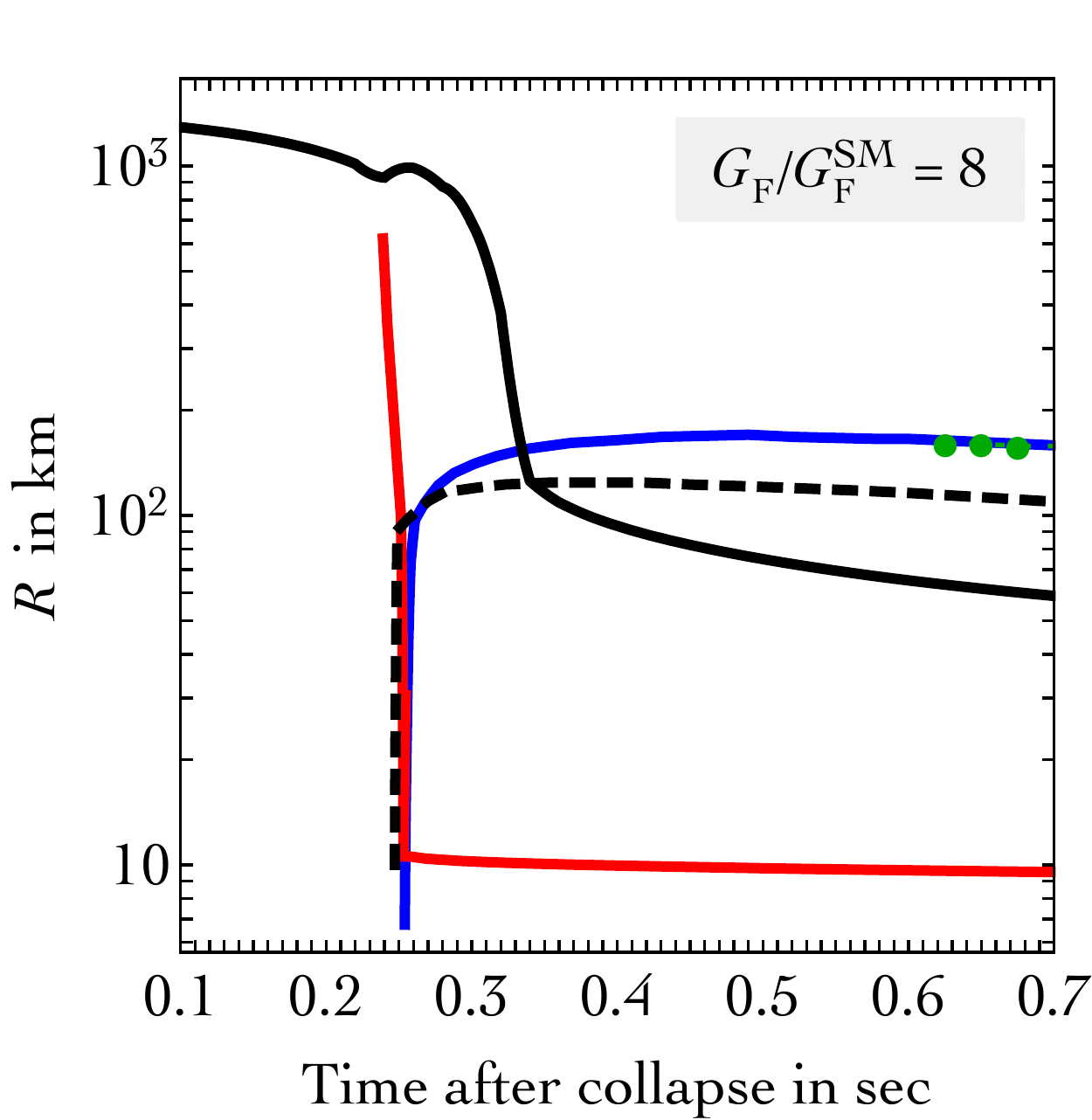}
\endminipage 
\caption{\em Time evolution of the relevant radii for $G_{\rm F}=\{1/4,1/2,1,2,4,8\}G_{\rm F}^{\rm SM}$.
We plot $R_{\rm Fe}$ (radius of the iron core, in black),
$R_{\rm ic}$ (radius of the inner iron core, in red),
$R_{\rm shock}$ (radius of the shock wave, in blue),
$R_\nu$ (radius of the neutrino-sphere, black-dashed) and 
$R_{\rm gain}$ (above which neutrino heating wins over cooling, green-dotted).
We simulated the collapse of a star with total mass $15M_\odot$ running the code in~\cite{0912.2393}.
As a proxy for $R_{\rm Fe}$ we plot the radius that encloses the mass $M=1.4M_\odot$.
Before the bounce, the inner core radius $R_{\rm ic}$ delimits the region of subsonic collapse from the supersonic outer core;
after the bounce, it represents the compact inner region of the nascent neutron star (here defined as the radial distance at which the 
entropy per baryon equals 3~\cite{1411.7058}).
\label{fig:simul}}
\end{figure}

\begin{figure}[!htb!]
 \centering
  \includegraphics[width=.48\linewidth]{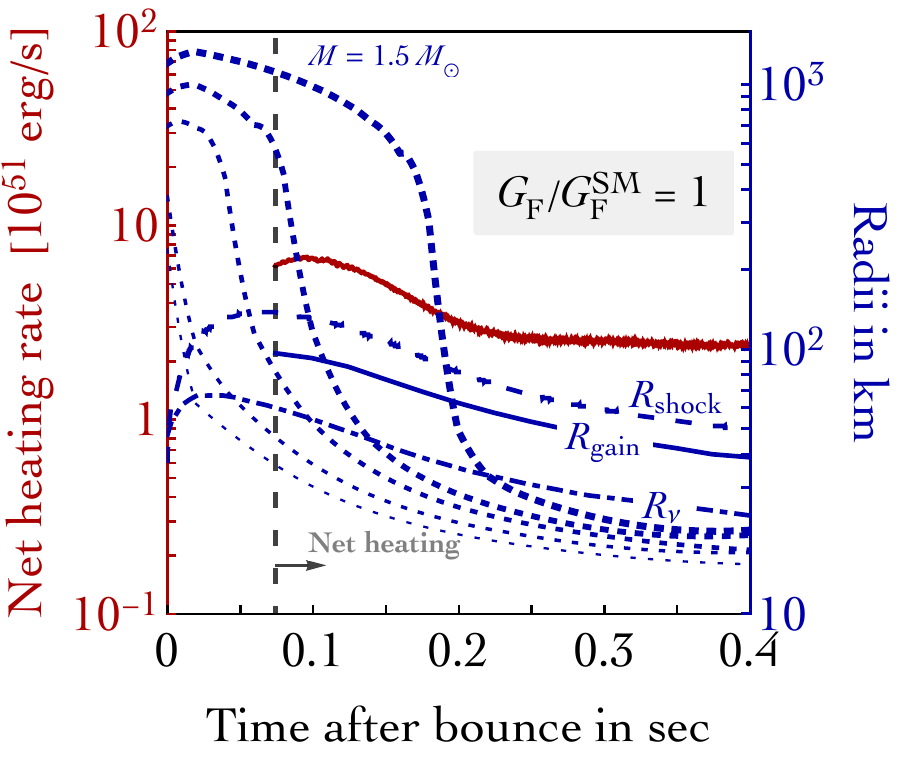}
\minipage{0.5\textwidth}\vspace{5mm}
\centering
  \includegraphics[width=.95\linewidth]{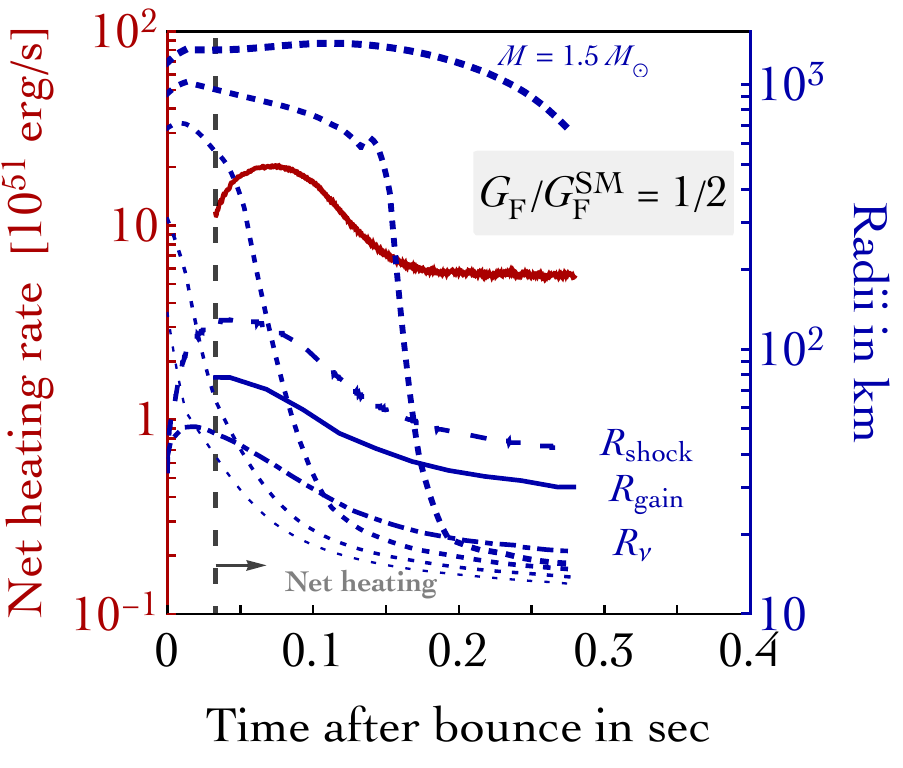}
\endminipage 
\minipage{0.5\textwidth}\vspace{5mm}
  \includegraphics[width=.95\linewidth]{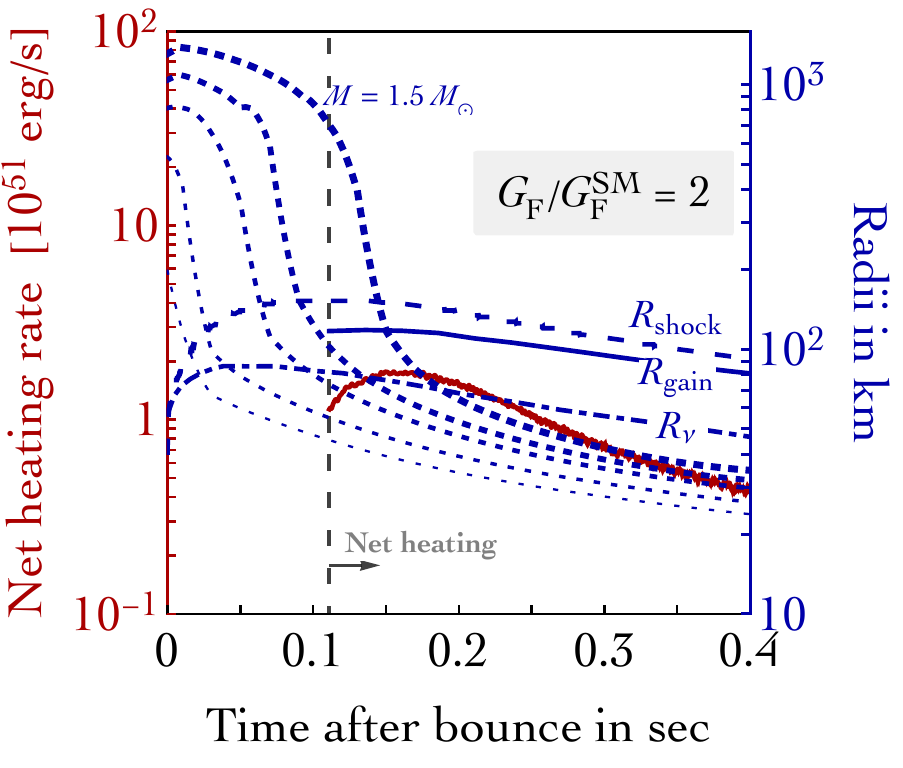}
\endminipage 
\caption{\em 
Time evolution of the total net heating rate for 
$\GF=\{{\color{red}1/2},1,{\color{blue}2}\} G_{\rm F}^{\rm SM}$ (left-side of the y-axes, in red).
We also show  (right-side of the y-axes, in blue) the gain radius and, again, the shock and neutrino-sphere radii.
Notice the qualitative scaling {\color{rossoc}{$Q_{\nu}$}}\,$\propto$\,{\color{navyblue}{$1/r^2$}} as in eq.\eq{Heat} neglecting cooling in $Q_{\nu}$.
The dotted blue curves show the contours of constant baryonic mass  (in progression, from thicker to thinner, $M = \{1.5,1.45,1.4,1.3,1.2\}\,M_{\odot}$:
the shock wave prevents the fall of the outer $\sim 15M_\odot$ 
during the simulated $0.4\,{\rm sec}$).
\label{fig:Comp}}
\end{figure}

\subsubsection*{Supernova explosions for smaller/larger Planck mass}

Given that the critical parameter is the dimension-less combination of eq.\eq{vcr1},
$M_{\rm Pl} m_n^3/v^4$, we expect that the range in $v$ argued in the previous section
becomes a strip in the ($v, M_{\rm Pl}$) plane, if both $v$ and $M_{\rm Pl}$ are varied.

Furthermore, and less importantly, the energies involved in the SN explosion scale
proportionally to $M_{\rm Pl}^3$.
These are
the gravitational energy
$E_{\rm tot}\sim M_{\rm Pl}^3/m_n^2$
and the comparable energy  needed to dissociate $N \sim (M_{\rm Pl}/m_n)^3$ iron nuclei
such that the shock wave can lead to a SN explosion.

However, we could not validate the above expectations trough reliable numerical simulations
at different values of the Planck mass.\footnote{Varying the Planck mass,
we only managed to run the numerical SN code of~\cite{0912.2393} 
in its version 1 (faster simplified treatment of neutrino energy transport)
and adapting progenitors computed at the physical value of the Planck mass.
Such simulations are possibly inadequate, so we do not report their results.}


%

\begin{figure}[t]
$$\includegraphics[width=0.48\textwidth]{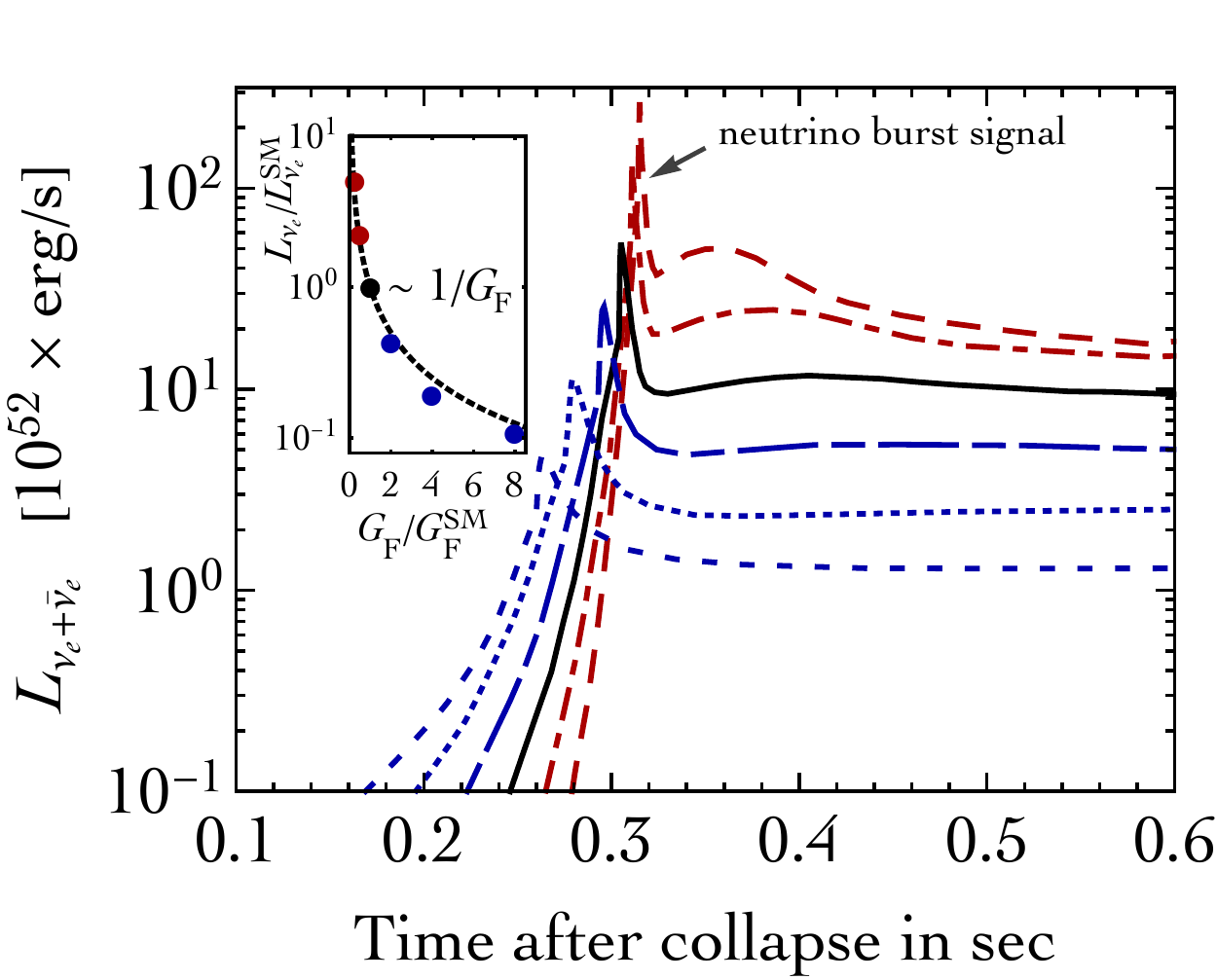}\qquad
\includegraphics[width=0.465\textwidth]{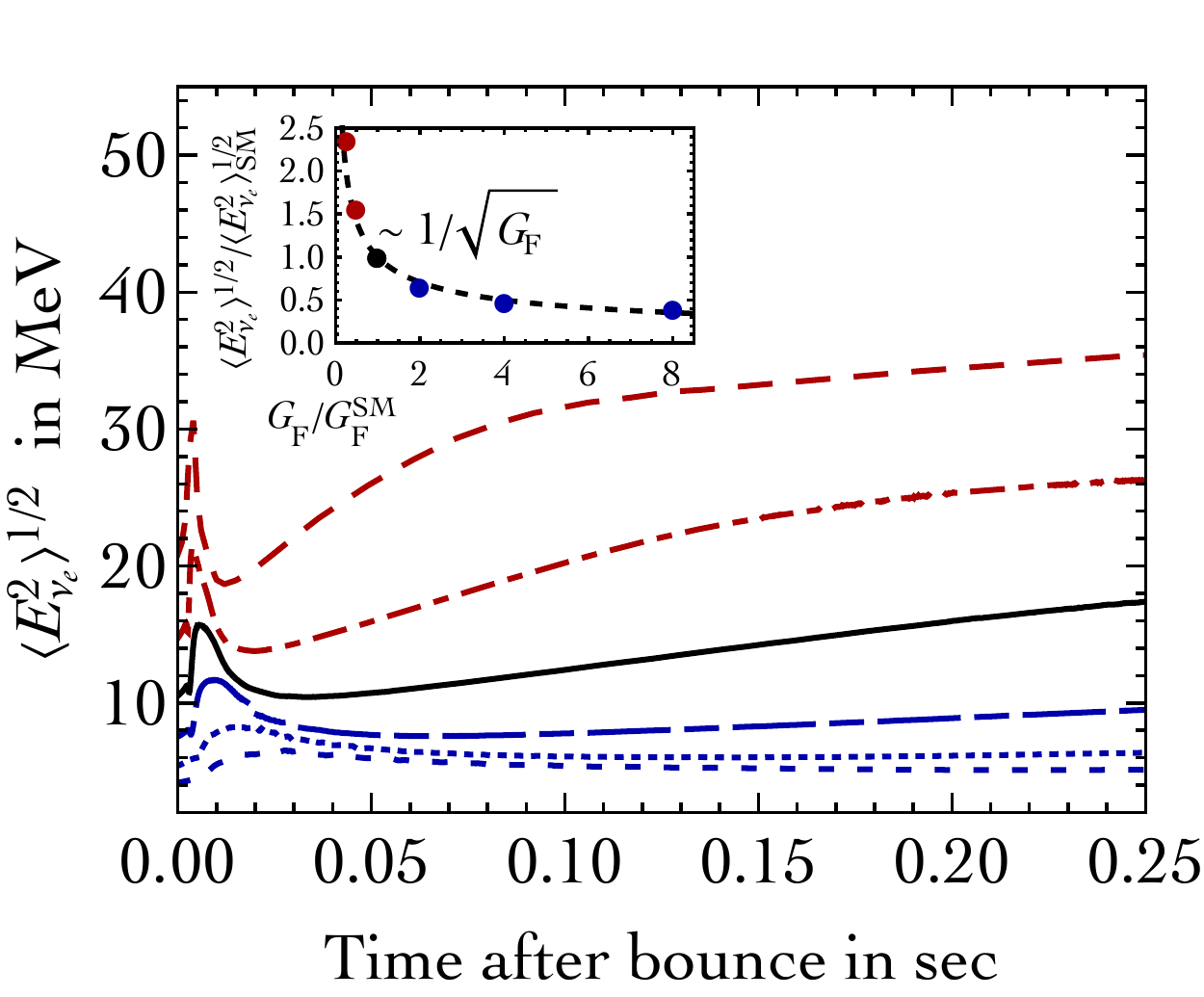}
$$
\caption{\em 
Time evolution of the power emitted in electron neutrinos (left)
and of their average quadratic energy (right)
for different values of the Fermi constant
$\GF=\{{\color{red}1/4,1/2},1,{\color{blue}2,4,8}\} G_{\rm F}^{\rm SM}$.
We simulated a SN with $M=15M_\odot$ running the code in~\cite{0912.2393}.
\label{fig:nu}}
\end{figure}

\subsection{Supernova explosions for different $v$: simulations}
Numerical simulations are needed to validate the above analytical discussion
because it involves not only simple rescaling of cross sections but also
disentangling dynamical adaptive features
(which result in non-trivial scaling laws) from numerical coincidences
(which make order one factors crucial).

We thereby run the public numerical SN code of~\cite{0912.2393} 
in its version 2, that involves an improved treatment of neutrino energy transport~\cite{1411.7058}.
We rescale all weak interactions changing the value of $v$ or equivalently of the Fermi constant
$\GF = 1/(2\sqrt{2} v^2)$.
The code employs the spherical approximation. 
Despite this simplification, simulations involve difficult numerical aspects and different time-scales:
the code has been optimised for the physical value $v=v_{\rm SM}$.
By changing $v$ optimisation gets lost, until numerical issues get out of control
(especially for larger $\GF \circa{>}\hbox{few}\times\GF^{\rm SM}$).
We thereby limit ourselves to run the code for relatively small deviations of $v$ from $v_{\rm SM}$, and
emphasize that the expertise 
of the authors of numerical SN codes seems needed for fully reliable simulations.
With this caveat in mind, we describe our results.

We start numerical simulations from a fixed star configuration with mass $M=15M_\odot$
and simulate the initial collapse --- a phase
nearly universal independently on the progenitor.
Weak interactions start playing a crucial role providing deleptonisation:
simulations indicate that $Y_e$ decreases in time 
triggering the collapse of the {\em inner} part of the core in roughly the usual way,
see fig.\fig{Ye}:
for smaller $\GF$ deleptonisation proceeds slower;
furthermore deleptonisation stops later because a larger density is needed to trap neutrinos:
as a result the final $Y_e$ becomes only slightly lower.
Thereby, for all simulated $v$, the
shock wave must cross the outer part of the core and can stall.

Fig.\fig{simul} shows the time evolution of the key radii discussed above,
for bigger (left) and smaller
(right) values of $v$.  
We find that, as expected, increasing $\GF$ increases 
$R_\nu \propto \GF^{1/2}$ (eq.\eq{RnuGF}).
Fig.\fig{Comp} also shows 
that the heating rate is smaller for large $\GF$ as expected 
in view of the larger gain radius $R_{\rm gain}\sim\hbox{few}\times R_\nu$.
Furthermore, fig.\fig{simul}  and \fig{Comp} show that 
the time-scale of neutrino cooling
grows for larger $\GF$, as expected.
Fig.\fig{nu} additionally shows that the luminosity of emitted neutrinos scales
as expected in eq.\eq{Lnu}, and that their average
 energy scales as expected in eq.\eq{RnuGF}.
Numerical simulations thereby confirm the expected scalings.

While our results indicate that supernova explosions need values
of the weak scale around its physical value,
we have not presented analytic estimates about the behaviour of the shock wave
around the physical value of $G_{\rm F}$, as this is a complicated issue 
and because non-sphericity seems anyhow needed for explosions. 
However, we can discuss results of numerical simulations 
that also compute how the above
changes in the dynamics of supernov\ae{} affect
the evolution of the shock radius.
The most evident result that emerges from fig.~\ref{fig:simul} is that the shock radius $R_{\rm shock}$ is smaller (bigger) for decreasing (increasing) values of $\GF$. 
To understand this behavior, it is instructive to consider the deleptonization of the material behind the shock. 
This is indeed an important physical quantity that keeps track of the evolution of the shock. 
The reason is the following. As the shock propagates, it loses energy by dissociating heavy nuclei into nucleons; in turn, this change of composition favors electron-capture processes 
 because the electron capture rate on free protons is significantly larger (because of its much smaller $Q$-value) than on neutron-rich elements. 
 Consequently, the outward-moving shock leaves behind, in its passing, a smaller value of the electron fraction $Y_e$.
We summarize our results in fig.~\ref{fig:YeMass} below. In the top panel we show the value of $Y_e$ as a function of the enclosed mass $M$ at different instants of time $t$ for the Standard Model value of $\GF$.
 The deleptonization effect is evident going from the pre-bounce phase (dashed lines) to the post-bounce phase (solid lines). In addition, the plot shows that 
 it is possible, in the post-bounce phase, to identify the sharp rise in $Y_e$ with the position of the shock front in terms of $M$. We checked that this alternative definition precisely matches the values of $R_{\rm shock}$ in our fig.~\ref{fig:simul}.
 We now turn to discuss the $\GF$-dependence of $R_{\rm shock}$. To gain some insight in this direction, we computed the matter pressure as a function of the radial distance in the post-bounce phase (with the corresponding time indicated with the subscript $t_{\rm pb}$).
 We show our results in the bottom panel of  fig.~\ref{fig:YeMass}. The sharp discontinuity in the radial profile of matter pressure indicates the position of the shock front. To facilitate the comparison among different values of $\GF$, we plot the matter pressure as a function of the radial distance in units of $R_{\rm shock}$.
 Right after the bounce at $t_{\rm pb} = 0.01\,{\rm s}$ (left panel), the radial profile of matter pressure does not significantly depend on $\GF$. 
 The situation changes if we consider, for instance,  $t_{\rm pb} = 0.25\,{\rm s}$ (right panel).  As relativistic electrons dominate the matter pressure in the region behind the shock, larger values of $\GF$ correspond to a more efficient electron-capture rate thus decreasing the matter pressure.  
 The pressure in the region beyond the shock is also modified (being related to the value behind the shock as a consequence of conservation of mass, momentum and energy  
 across the discontinuity of the shock front, as dictated by Rankine-Hugoniot conditions). 
 The increased value of pressure makes the pressure gradient more negative, and, consequently,  it pushes the shock inwards.
To justify this point, remember that in General Relativity for a perfect fluid with mass-energy density $\varepsilon(r)$ and pressure $P(r)$, the  Tolman-Oppenheimer-Volkoff equation reads
\begin{align}\label{eq:TOV}
\frac{dP(r)}{dr} = -\frac{G_N[P(r) + \varepsilon(r)]}{r[r-2G_N m(r)]}\left[
m(r) + 4\pi r^3 P(r)
\right]\,,
\end{align}
 with $G_N$ the Newton's constant and $m(r)$ representing the mass-energy inside the shell of radial coordinate $r$. Although written in a stationary approximation that is not suitable for our purposes, 
the right-hand side of  eq.\,(\ref{eq:TOV}) shows that pressure gravitates exactly like the mass-energy distribution, thus contributing to the inward-directed pull of gravity.

\medskip

\begin{figure}[!h!]
 \centering
  \includegraphics[width=.48\linewidth]{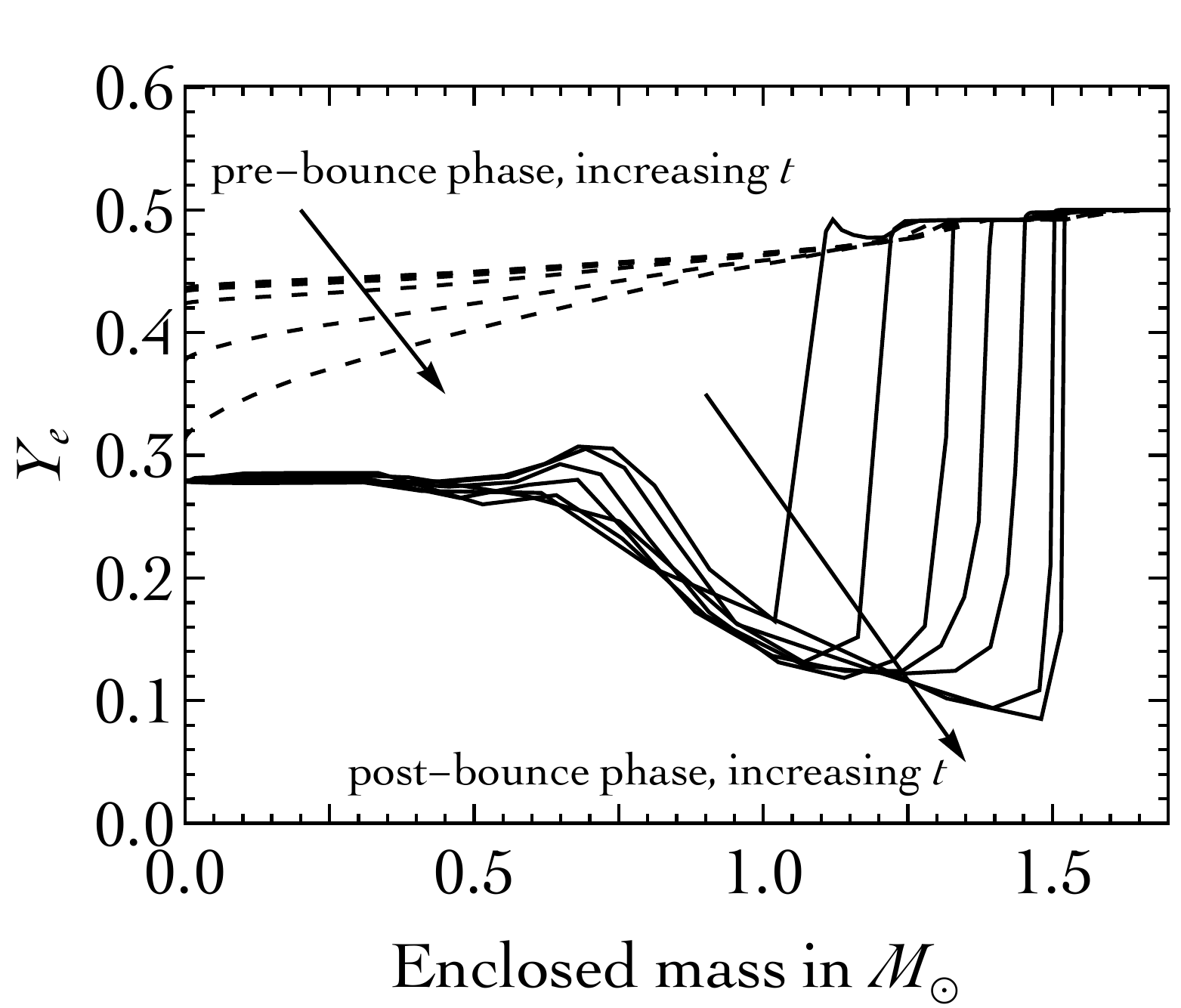}
\minipage{0.5\textwidth}\vspace{5mm}
\centering
  \includegraphics[width=.95\linewidth]{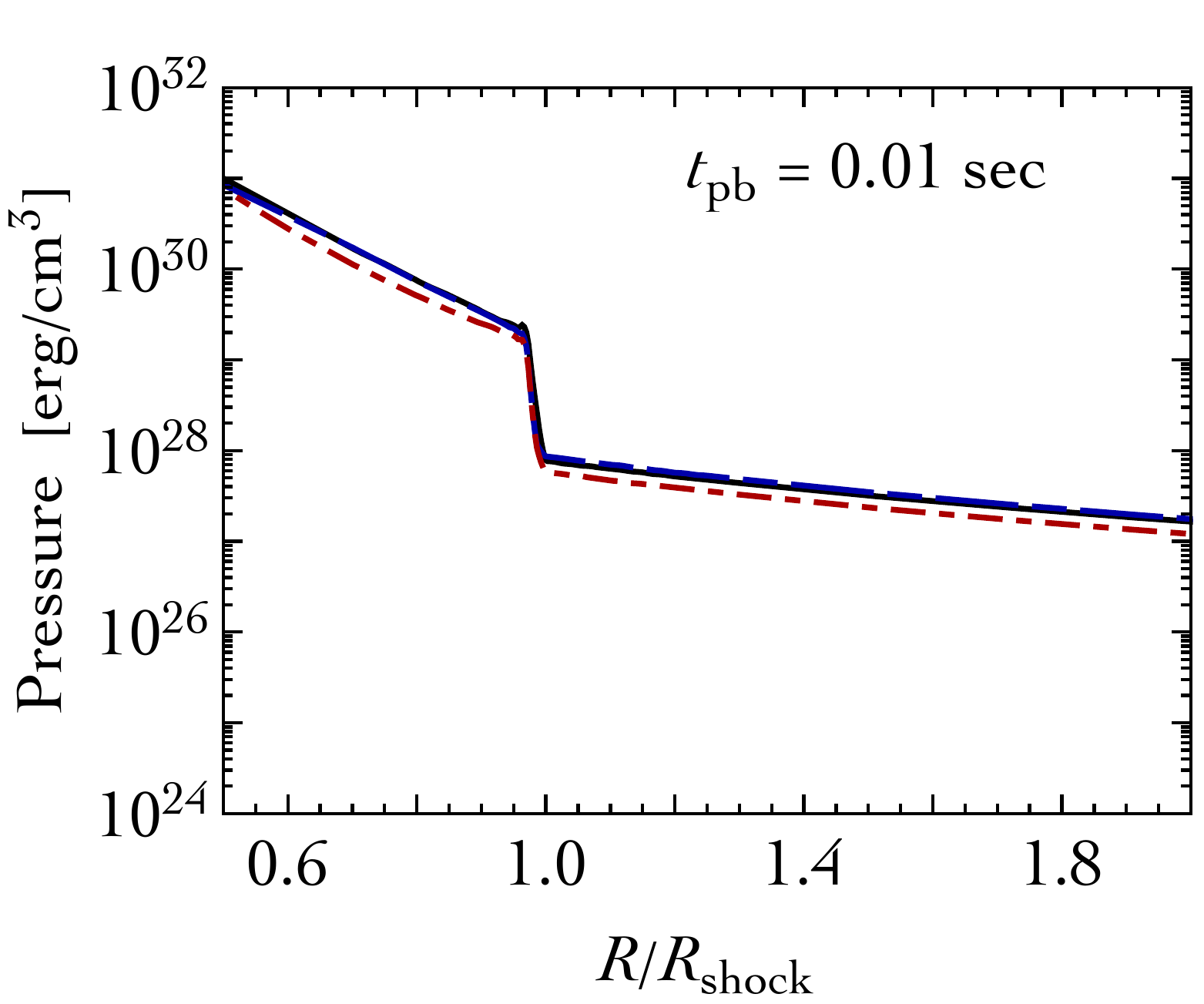}
\endminipage 
\minipage{0.5\textwidth}\vspace{5mm}
  \includegraphics[width=.95\linewidth]{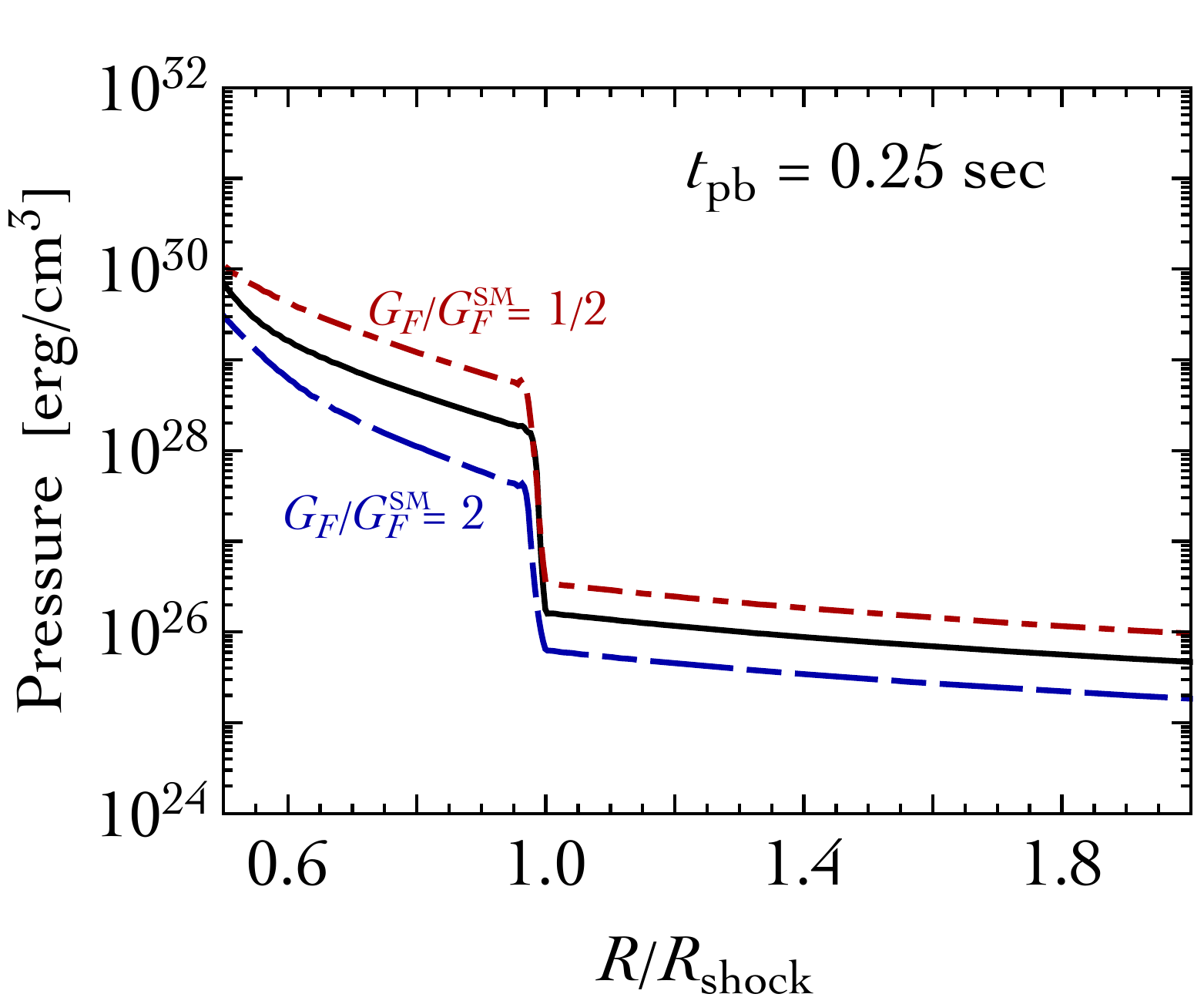}
\endminipage 
\caption{\em 
{\bf Top:} Electron fraction as a function of the enclosed mass for $\GF = \GF^{\rm SM}$ and for different instants of time with
 dashed (solid) lines referring to the pre(post)-bounce phase.
{\bf Bottom:} Matter pressure as a function of the radial distance $R$ (in units of $R_{\rm shock}$) for  different values of $\GF$ ($\GF = \GF^{\rm SM}/2$ and $\GF = 2\GF^{\rm SM}$, see labels) evaluated at post-bounce time $t_{\rm pb} = 0.01\,{\rm s}$ (left panel) and $t_{\rm pb} = 0.25\,{\rm s}$ (right panel).
\label{fig:YeMass}}
\end{figure}
 Finally, we see from fig.\fig{simul} and \fig{Comp} 
that for $v \sim v_{\rm SM}$
 $R_{\rm shock}$ manages to reach the gain region,
 $R_{\rm shock}\circa{>} R_{\rm gain}\propto \GF^{1/2}$.
Furthermore, the
numerical code of~\cite{0912.2393}  stops working at values of $\GF$ so large that
the neutrino-sphere becomes comparable to the stalled
shock wave: simulating point 4 above would need a dedicated code. 
The numerical simulations of fig.\fig{simul} confirm 
that the critical $v_{\rm trap}$ is a factor of few above the physical value $v_{\rm SM}=174\GeV$,
but cannot determine it precisely.


\medskip

Indeed, we never get any explosion in spherical approximation~\cite{astro-ph/0006418}.
As well known from simulations at the physical value of the weak scale,
explosions seems to need 2d~\cite{astro-ph/0312633,astro-ph/0507135,1212.1747,1409.5779} or 3d~\cite{1505.05110,1506.05139,1807.07579,1809.05106} simulations, 
that are computationally much more intensive
than simulations in spherical approximation.
We hope that experts in supernova physics can test our main findings:
that neutrino-driven SN explosions happen in a restricted range of the weak scale
$v$ that contains its physical value $v_{\rm SM}$.
The upper bound of $v$ is especially important for fundamental theory,
given that it seems to have anthropic relevance.
Furthermore, once explosions are simulated,
it would be interesting to compute the fraction of the total energy that explodes into nuclei
as function of $v$.
At the physical $v_{\rm SM}$ this fraction is about $1\%$:
SN explosions spread the elements needed for life, and also
damage life in the nearby $\sim$30 ly
(in the solar neighbourhood this corresponds to a rate comparable to the time span of life, about $0.5$ Gyr).

\section{Stellar evolution}\label{star}

A too small $v$ is anthropically excluded because ordinary
matter at temperature $T$ would cool too fast 
loosing energy into neutrinos with mass $m_\nu \circa{<} T$, 
with a time scale
$ \tau_{\rm cool} \sim    \sfrac{v^4}{\alpha^3(m_e T)^{3/2}}$.
Complex chemistry and `life' is possible at the `ambient' temperature comparable to the binding
energy of atoms, $T \sim  \alpha^2 m_e$.
At this temperature $\tau_{\rm cool}  \sim  \sfrac{v^4}{\alpha^8 m_e^5}$
is much larger than the age of the Universe.
Furthermore, matter is heated by stars.
Since stars have higher temperature, 
weak interactions play a role in stellar evolution.
A non-standard value of $v$ would modify stellar evolution in two ways: by changing
\begin{itemize}
\item[1.]  energy losses into neutrinos, and
\item[2.] weak interactions that contribute to star burning.
\end{itemize}

The second factor is not crucial, because the  cross sections that
depend on the weak scale, also depend more strongly 
on the Coloumb barrier factor $e^{-3 E_{\rm Gamow}(T)/T}$
such that a steady state of stellar burning
is restored by a small change in $T$.
Even in the extreme weak-less limit, stars can anyhow burn through purely
nuclear interactions which do not involve weak interactions.
In particular, BBN at large $v$ leads to an equal number of neutrons and protons
i.e.\ to Helium;
Helium nuclei can burn to $^{12}_{~6}{\rm C}$ through the triple
$\alpha$ process mediated by the Hoyle resonance.

\smallskip

The first factor depends strongly on the stellar temperature and density.
Considering, for example, the Sun, only a small ($10^{-10}$) fraction of its energy is emitted as
thermal radiation of neutrinos with energy comparable to its central 
temperature $T =1.3\keV$~\cite{1708.02248}.\footnote{The sun  emits another $\approx 3\%$ of its energy in MeV neutrinos, because the same nuclear/weak interactions
that produce the solar energy (in particular $pp \to d \bar e\nu_e$) also emit neutrinos.
Such neutrinos cannot be considered as energy loss.}
Since the thermal neutrino rate is proportional to $1/v^4$, 
a small $v\circa{<} v_{\rm SM} /300$ would modify the solar behaviour.
Different processes ($eZ\to e Z\nu\bar\nu$,
$e\gamma \to e\nu\bar\nu$, pair production,
plasmon decay $\gamma \to\nu\bar\nu$)
dominate energy losses into neutrinos for different stellar temperatures and densities, 
and their rates grow with the temperature proportionally to $T^{3-9}$, depending on the process.
For the physical value of $v$, neutrino radiation dominates energy losses of stars with central temperatures hotter than $T \circa{>}50\keV$, making the later stages of stellar evolution very fast.

\smallskip

The stellar temperature is determined by the stellar mass (and chemical composition)
as dictated by stellar evolution.
All stars have masses around $M_{\rm Pl}^3/m_p^2$ due to physics which does not involve weak nor strong interactions~\cite{Carr:1979sg,Adams:2008ad}.
More precisely, star masses $M$ must lie in a range $M_{\rm min} < M < M_{\rm max}$.
The minimal stellar mass $M_{\rm min} \sim (T_{\rm nuc}/m_e)^{3/4}  M_{\rm Pl}^3/m_p^2$
arises because the star must reach the critical temperature $T_{\rm nuc} \sim \alpha^2 m_n$ 
before that nuclear reactions proceed igniting the star.
The maximal stellar mass $M_{\rm max}\sim M_{\rm Pl}^3/m_p^2$
arises because radiation pressure $p \sim T^4$ dominates if $T$ is too large, making stars unstable.
Precise computations find
$M_{\rm min} \approx 0.08 M_\odot$ and $M_{\rm max}\approx 100 M_\odot$
at the physical values of the fundamental constants~\cite{Adams:2008ad}.
Varying $M_{\rm Pl}$ only, order one factors change and
the range closes ($M_{\rm min}=M_{\rm max}$) and stars disappear if
$M_{\rm Pl} \circa{<} M_{\rm Pl}^{\rm SM}/100$~\cite{Adams:2008ad}.
A stronger limit $M_{\rm Pl} \circa{<} M_{\rm Pl}^{\rm SM}/12$ arises demanding
that stars are long-lived enough~\cite{1511.06958}.
 
We studied how stellar evolution changes for different values of $v$,
finding that for wide ranges of $v$ stars settle to different steady-state regimes,
which can be slow enough to support `life' as well as fast enough to produce the first 
nuclei.\footnote{Stellar
evolution has been studied in collaboration with Giada Valle, Matteo Dell'Omodarme, Scilla Degl'Innocenti and Pier Giorgio Prada Moroni.}
We do not document our findings, as the
same conclusion has been recently reached in~\cite{1809.05128}.
While stellar evolution depends on $v$
(such that the weak constant could be measured from stellar data), 
no anthropic boundary on $v$ is found from stellar evolution.
Presumably the only anthropic boundary is
$M_{\rm Pl} \circa{>} M_{\rm Pl}^{\rm SM}/100$
if both $v$ and $M_{\rm Pl}$ are varied.

\section{Conclusion and discussion}\label{concl}
Anthropic arguments,
despite their controversial reputation, are important for indicating main directions in
fundamental physics.
We explored whether anthropic selection played a role in selecting the value of the weak scale $v$.
As discussed in the introduction, fermion masses $m_f = y_f v$ are anthropically relevant, but they depend on Yukawa couplings $y_f$ times $v$, so that they do not directly restrict $v$,
leading to a paradoxical situation.

We focused on physical processes  {\em directly} affected by weak interactions:
the ones of possible anthropic relevance are BBN, stellar evolution and core-collapse supernov\ae{}.
We studied what changes if the Higgs vacuum expectation value $v$
(and thereby the Fermi constant $\GF= 1/(2\sqrt{2} v^2)$  that controls weak interactions)
differs from its physical value, with the quark and lepton masses kept fixed.

BBN was studied in~\cite{1409.0551} and stellar evolution in~\cite{1809.05128}:
they do not seem to lead to anthropic boundaries on $v$.
Indeed, stellar evolution changes qualitatively if $v$ is changed by more than one order in magnitude in either direction, but stars still burn in a slow stable way.
For  large $v$ weak nuclear processes (such as $pp$ burning) no longer lead to stable stars,
which anyhow continue existing thanks to strong nuclear processes (such as the triple $\alpha$ process).
Stellar dynamics seem to lead to a weak anthropic bound on the Planck mass: it must be
larger than  $1\%$ of its observed value~\cite{Adams:2008ad}.

\begin{figure}[t]
\begin{center}
\includegraphics[width=0.95\textwidth]{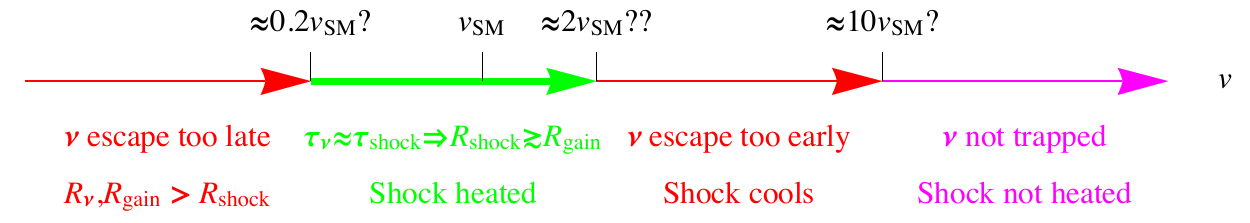}
\caption{\em Behaviour of core-collapse supernov\ae{}
as function of the weak scale $v$.
Although numerical factors are only indicative,
we conclude that neutrino-driven SN
explosions arise in a narrow range of $v$.
\label{fig:SNvsummarybarra}}
\end{center}
\end{figure}

\medskip

The situation with supernov\ae{} (SN) seems more interesting, as
core-collapse supernova explosions seem anthropically relevant and driven by weak interactions.
We assume 
\begin{itemize}
\item the dominant (but not necessarily correct) paradigm according to which
core-collapse SN explosions happen thanks to weak interactions: 
neutrinos push the stalled shock wave
generated by the rebounce of the inner core when the supernova reaches nuclear density,
such that the rejuvenated shock manages to spread the outer SN material.

\item that the observed core-collapse SN explosions are anthropically relevant,
given that they are the largely dominant process that spreads elements possibly needed for `life',
in particular oxygen (see fig.\fig{PeriodicTable}).
\end{itemize}
The two points above are plausible, but establishing them is difficult.
Just to be very clear, we repeat that our
subsequent discussion relies on the assumption
that light elements  needed for `life' are significantly produced only by neutrino-driven explosions of
core-collapse supernov\ae.


\medskip

We argued that core-collapse supernov\ae{} no longer explode if $v$ is increased 
or decreased by a factor of few,
as illustrated in fig.\fig{SNvsummarybarra}.
For the physical value of $v$, the time-scale of neutrino trapping
matches the gravitational time-scale of the supernova
(with a related matching of the spatial scales).
\begin{itemize}
\item Increasing $v$ reduces weak interactions, such that neutrinos 
escape too fast and no longer push the shock wave  when it stalls.

Ultimately, at large $v \circa{>}10 v_{\rm SM}$  neutrinos become not trapped.

\item Decreasing $v$ increases weak interactions, such that neutrinos
exit too late for rejuvenating the shock wave before the collapse of the exterior material.

Ultimately, at small $v \circa{<}0.2 v_{\rm SM}$
 the gain-sphere
(the region where neutrinos push matter) and the neutrino-sphere
(which is the region where neutrinos are trapped) grow bigger than the 
shock wave, such that neutrinos no longer push the shock outwards.
\end{itemize}
We provided analytic estimates that capture supernova physics,
disentangling adaptive dynamical features from accidental numerical coincidences.
For example, the  energy transmitted by interactions of
trapped neutrinos is comparable
to the total energy, independently of the value of $v$.
Disentangling dynamics from tunings is needed to correctly identify
anthropic features~\cite{1101.2547}.
We validated aspects of our analytic understanding relying on numerical simulations
in spherical approximation.
However this approximation does not lead to supernova explosions.
As well known, supernova explosions at $v=v_{\rm SM}$ seem so much critical
that non-spherical simulations are needed to get 
enhancements by a few $10\%$ which lead to explosions.
We hope that dedicated work by experts can firmly establish (or revise) our results.

\medskip

This finding has important implications for fundamental physics.
As hinted by our sub-title,
like most theorists we would have preferred an understanding of the weak scale
based on natural super-symmetry rather than on anthropic super-nov\ae{}.
Paraphrasing Bohr, 
anthropic arguments work even when physicists don't believe in them~\cite{Bohr}.
As discussed in the introduction,
our direct anthropic boundary on the weak scale (if confirmed by future studies)
avoids the paradox raised by previous anthropic bounds:
on fermion masses $m_f = y_f v$:
a SM-like theory with fixed $m_f$ and bigger $v$ would need a less unlikely tuning of $v^2/M_{\rm Pl}^2$.


\footnotesize
\subsubsection*{Acknowledgements}
This work was supported by the ERC grant NEO-NAT. 
The work of A.U. was partly supported by the INFN grant SESAMO.
We thank Kfir Blum, Evan O'Connor, Scilla Degl'Innocenti,
Matteo Dell'Omodarme,
Marco Nardecchia, Giorgio Prada Moroni,
Christian Ott, Thomas Janka, Giada Valle for useful discussions.
A.U. thanks Kavli IPMU, where  this  project  was  completed, and the city of Tokyo for  the  warm and kind  hospitality.


\begin{thebibliography}{nnn}\bibitem{Weinberg:1988cp}
\article[Weinberg:1988cp]{S. Weinberg}{Rev. Mod. Phys.}{61}{1}{1989}
{The Cosmological Constant Problem}.


\bibitem{hep-th/0603249}
\heparticle[hep-th/0603249]{J. Polchinski}{The Cosmological Constant and the String Landscape}.


\bibitem{1407.2122}
\article[1407.2122]{G. Altarelli}{Frascati Phys. Ser.}{58}{102}{2014}
{The Higgs and the Excessive Success of the Standard Model}.


\bibitem{Wein}
\article[Weinberg:1987dv]{S. Weinberg}{Phys. Rev. Lett.}{59}{2607}{1987}
{Anthropic Bound on the Cosmological Constant}.


\bibitem{SmallLambda}
See e.g.\
\article[1801.08781]{L.A. Barnes, P.J. Elahi, J. Salcido, R.G. Bower, G.F. Lewis, T. Theuns, M. Schaller, R.A. Crain, J. Schaye}{Mon. Not. Roy. Astron. Soc.}{477}{3727}{2018}
{Galaxy Formation Efficiency and the Multiverse Explanation of the Cosmological Constant with EAGLE Simulations}.


\bibitem{hep-th/0004134}
\article[hep-th/0004134]{R. Bousso, J. Polchinski}{JHEP}{0006}{006}{2000}
{Quantization of four form fluxes and dynamical neutralization of the cosmological constant}.


\bibitem{hep-th/0302219}
\heparticle[hep-th/0302219]{L. Susskind}{The Anthropic landscape of string theory}.


\bibitem{hep-th/0301240}
\article[hep-th/0301240]{S. Kachru, R. Kallosh, A.D. Linde, S.P. Trivedi}{Phys. Rev.}{D68}{046005}{2003}
{De Sitter vacua in string theory}.


\bibitem{hep-ph/9707380}
\article[hep-ph/9707380]{V. Agrawal, S.M. Barr, J.F. Donoghue, D. Seckel}{Phys. Rev.}{D57}{5480}{1997}
{The Anthropic principle and the mass scale of the standard model}.


\bibitem{0712.2454}
\article[0712.2454]{L.J. Hall, Y. Nomura}{Phys. Rev.}{D78}{035001}{2007}
{Evidence for the Multiverse in the Standard Model and Beyond}.


\bibitem{hep-ph/0703219}
\article[hep-ph/0703219]{S.M. Barr, A. Khan}{Phys. Rev.}{D76}{045002}{2007}
{Anthropic tuning of the weak scale and of $m_u/m_d$ in two-Higgs-doublet models}.


\bibitem{0809.1647}
\article[0809.1647]{R.L. Jaffe, A. Jenkins, I. Kimchi}{Phys. Rev.}{D79}{065014}{2008}
{Quark Masses: An Environmental Impact Statement}.


\bibitem{hep-ph/9811386}
\article[hep-ph/9811386]{L. Giusti, A. Romanino, A. Strumia}{Nucl. Phys.}{B550}{3}{1998}
{Natural ranges of supersymmetric signals}.


\bibitem{hep-ph/0007265}
\heparticle[hep-ph/0007265]{R. Barbieri, A. Strumia}{The 'LEP paradox'}.


\bibitem{1607.01659}
\article[1607.01659]{F. Sannino, A. Strumia, A. Tesi, E. Vigiani}{JHEP}{1611}{029}{2016}
{Fundamental partial compositeness}.


\bibitem{1303.7244}
\article[1303.7244]{M. Farina, D. Pappadopulo, A. Strumia}{JHEP}{1308}{022}{2013}
{A modified naturalness principle and its experimental tests}.


\bibitem{1504.05403}
\article[1504.05403]{M. Fabbrichesi, A. Urbano}{Phys. Rev.}{D92}{015028}{2015}
{Naturalness redux: The case of the neutrino seesaw mechanism}.


\bibitem{1705.03896}
\article[1705.03896]{A. Salvio, A. Strumia}{Eur. Phys. J.}{C78}{124}{2018}
{Agravity up to infinite energy}.


\bibitem{Carter:1974zz}
\article[Carter:1974zz]{B. Carter}{IAU Symp.}{63}{291}{1974}
{Large number coincidences and the anthropic principle in cosmology}.


\bibitem{Carr:1979sg}
\article[Carr:1979sg]{B.J. Carr, M.J. Rees}{Nature}{278}{605}{1979}
{The anthropic principle and the structure of the physical world}.


\bibitem{Barrow:1988yia}
\heparticle[Barrow:1988yia]{J.D. Barrow, F.J. Tipler}{The Anthropic Cosmological Principle}.


\bibitem{1409.0551}
\article[1409.0551]{L.J. Hall, D. Pinner, J.T. Ruderman}{JHEP}{1412}{134}{2014}
{The Weak Scale from BBN}.


\bibitem{astro-ph/0108071}
\article[astro-ph/0108071]{S.P. Oh, Z. Haiman}{Astrophys. J.}{569}{558}{2001}
{Second-generation objects in the universe: radiative cooling and collapse of halos with virial temperatures above $10^4$ kelvin}.


\bibitem{Astroper}
\article{J.A. Johnson}{Science}{263}{474}{2019}
{\href{https://science.sciencemag.org/content/363/6426/474/tab-pdf}{Populating the periodic table: Nucleosynthesis of the elements}}.


\bibitem{Woosley:1995ip}
\article[Woosley:1995ip]{S.E. Woosley, T.A. Weaver}{Astrophys. J. Suppl.}{101}{181}{1995}
{The Evolution and explosion of massive stars. 2. Explosive hydrodynamics and nucleosynthesis}.


\bibitem{hep-ph/0609050}
\heparticle[hep-ph/0609050]{L. Clavelli, III. White}{Problems in a weakless universe}.


\bibitem{1506.02655}
\heparticle[1506.02655]{D. Kushnir}{The progenitors of core-collapse supernovae suggest thermonuclear origin for the explosions}.


\bibitem{1601.03422}
\article[1601.03422]{K. Blum, D. Kushnir}{Astrophys. J.}{828}{31}{2016}
{Neutrino Signal of Collapse-induced Thermonuclear Supernovae: the Case for Prompt Black Hole Formation in SN1987A}.


\bibitem{1811.11178}
\heparticle[1811.11178]{N. Bar, K. Blum, G. D'Amico}{Neutrino flavour as a test of the explosion mechanism of core-collapse supernovae}.


\bibitem{hep-ph/0606054}
\heparticle[hep-ph/0606054]{A. Strumia, F. Vissani}{Neutrino masses and mixings and...}.


\bibitem{astro-ph/0612072}
\article[astro-ph/0612072]{H-T. Janka, K. Langanke, A. Marek, G. Martinez-Pinedo, B. Mueller}{Phys. Rept.}{442}{38}{2006}
{Theory of Core-Collapse Supernovae}.


\bibitem{1206.2503}
\article[1206.2503]{H-T. Janka}{Ann. Rev. Nucl. Part. Sci.}{62}{407}{2012}
{Explosion Mechanisms of Core-Collapse Supernovae}.


\bibitem{1702.08713}
\heparticle[1702.08713]{H.-T. Janka}{Neutrino Emission from Supernovae}.


\bibitem{astro-ph/0211194}
\article[astro-ph/0211194]{T.A. Thompson, A. Burrows, P.A. Pinto}{Astrophys. J.}{592}{434}{2002}
{Shock breakout in core-collapse supernovae and its neutrino signature}.


\bibitem{Colgate:1966ax}
\article[Colgate:1966ax]{S.A. Colgate, R.H. White}{Astrophys. J.}{143}{626}{1966}
{The Hydrodynamic Behavior of Supernovae Explosions}.


\bibitem{Bethe:1984ux}
\article[Bethe:1984ux]{H.A. Bethe, J.R. Wilson}{Astrophys. J.}{295}{14}{1985}
{Revival of a stalled supernova shock by neutrino heating}.


\bibitem{astro-ph/0006418}
\article[astro-ph/0006418]{M. Liebendoerfer, A. Mezzacappa, F-K. Thielemann, O.E.B. Messer, W.R. Hix, S.W. Bruenn}{Phys. Rev.}{D63}{103004}{2000}
{Probing the gravitational well: no supernova explosion in spherical symmetry with general relativistic boltzmann neutrino transport}.


\bibitem{astro-ph/0312633}
\article[astro-ph/0312633]{E. Livne, A. Burrows, R. Walder, I. Lichtenstadt, T.A. Thompson}{Astrophys. J.}{609}{277}{2003}
{Two - dimensional, time - dependent, multi-group, multi-angle radiation hydrodynamics test simulation in the core-collapse supernova context}.


\bibitem{astro-ph/0507135}
\article[astro-ph/0507135]{R. Buras, M. Rampp, H.-T. Janka, K. Kifonidis}{Astron. Astrophys.}{447}{1049}{2005}
{Two-dimensional hydrodynamic core-collapse supernova simulations with spectral neutrino transport. 1. Numerical method and results for a 15 solar mass star}.


\bibitem{1212.1747}
\article[1212.1747]{S.W. Bruenn, A. Mezzacappa, W.R. Hix, E.J. Lentz, O.E.B. Messer, E.J. Lingerfelt, J.M. Blondin, E. Endeve, P. Marronetti, K.N. Yakunin}{Astrophys. J.}{767}{L6}{2012}
{Axisymmetric Ab Initio Core-Collapse Supernova Simulations of $12$-$25 M_\odot$ Stars}.


\bibitem{1409.5779}
\article[1409.5779]{
S.W. Bruenn et al.}{Astrophys. J.}{818}{123}{2016}
{The Development of Explosions in Axisymmetric ab initio Core-Collapse Supernova Simulations of 12-25 $M_\odot$ Stars}.


\bibitem{1505.05110}
\article[1505.05110]{E.J. Lentz, S.W. Bruenn, W.R. Hix, A. Mezzacappa, O.E.B. Messer, E. Endeve, J.M. Blondin, J.A. Harris, P. Marronetti, K.N. Yakunin}{Astrophys. J.}{807}{L31}{2015}
{Three-dimensional Core-collapse Supernova Simulated Using a $15 M_\odot$ Progenitor}.


\bibitem{1506.05139}
\article[1506.05139]{B. M{\" u}ller}{Mon. Not. Roy. Astron. Soc.}{453}{287}{2015}
{The Dynamics of Neutrino-Driven Supernova Explosions after Shock Revival in 2D and 3D}.


\bibitem{1807.07579}
\article[1807.07579]{E.P. O'Connor, S.M. Couch}{Astrophys. J.}{865}{81}{2018}
{Exploring Fundamentally Three-dimensional Phenomena in High-fidelity Simulations of Core-collapse Supernovae}.


\bibitem{1809.05106}
\article[1809.05106]{D. Vartanyan, A. Burrows, D. Radice, A.M. Skinner, J. Dolence}{Mon. Not. Roy. Astron. Soc.}{482}{351}{2019}
{A Successful 3D Core-Collapse Supernova Explosion Model}.


\bibitem{hep-ph/0604027}
\article[hep-ph/0604027]{R. Harnik, G.D. Kribs, G. Perez}{Phys. Rev.}{D74}{035006}{2006}
{A Universe without weak interactions}.


\bibitem{1010.5550}
\article[1010.5550]{E. O'Connor, C.D. Ott}{Astrophys. J.}{730}{70}{2010}
{Black Hole Formation in Failing Core-Collapse Supernovae}.


\bibitem{1501.02845}
\article[1501.02845]{A. Perego, M. Hempel, C. Fr{\" o}hlich, K. Ebinger, M. Eichler, J. Casanova, M. Liebendoerfer, F.-K. Thielemann}{Astrophys. J.}{806}{275}{2015}
{Pushing Core-collapse Supernovae to Explosions in Spherical Symmetry. i. the Model and the Case of sn 1987a}.


\bibitem{0912.2393}
\article[0912.2393]{E. O'Connor, C.D. Ott}{Class. Quant. Grav.}{27}{114103}{2009}
{A New Open-Source Code for Spherically-Symmetric Stellar Collapse to Neutron Stars and Black Holes}.


\bibitem{1411.7058}
\article[1411.7058]{E. O'Connor}{Astrophys. J. Suppl.}{219}{24}{2015}
{An Open-Source Neutrino Radiation Hydrodynamics Code for Core-Collapse Supernovae}.


\bibitem{1708.02248}
\article[1708.02248]{E. Vitagliano, J. Redondo, G. Raffelt}{JCAP}{1712}{010}{2017}
{Solar neutrino flux at keV energies}.


\bibitem{Adams:2008ad}
\article[0807.3697]{F.C. Adams}{JCAP}{0808}{010}{2008}
{Stars In Other Universes: Stellar structure with different fundamental constants}.


\bibitem{1511.06958}
\article[1511.06958]{F.C. Adams}{JCAP}{1602}{042}{2016}
{Constraints on Alternate Universes: Stars and habitable planets with different fundamental constants}.


\bibitem{1809.05128}
\article[1809.05128]{A.R. Howe, E. Grohs, F.C. Adams}{Phys. Rev.}{D98}{063014}{2018}
{Nuclear Processes in Other Universes: Varying the Strength of the Weak Force}.


\bibitem{Petrosian:1967alk}
\article[Petrosian:1967alk]{V. Petrosian, G. Beaudet, E.E. Salpeter}{Phys. Rev.}{154}{1445}{1967}
{Photoneutrino Energy Loss Rates}.


\bibitem{1101.2547}
\article[1101.2547]{E. Epelbaum, H. Krebs, D. Lee, U-G. Meissner}{Phys. Rev. Lett.}{106}{192501}{2011}
{Ab initio calculation of the Hoyle state}.


\bibitem{Bohr}
N. Bohr, private communication about his horseshoe.


\end{thebibliography}
\end{document}